\documentclass[article]{IEEEtran}
\IEEEoverridecommandlockouts

\usepackage{graphicx}
\usepackage{amsmath,amssymb}
\usepackage{cases}
\usepackage{color}
\usepackage{amsfonts}
\usepackage{indentfirst}
\usepackage{float}
\usepackage{setspace}
\usepackage{cite}
\usepackage{caption}
\usepackage{algorithm}
\usepackage{algpseudocode}
\usepackage{booktabs}
\usepackage{subfigure}
\usepackage{multirow}
\usepackage{amsthm}
\usepackage{epstopdf}
\usepackage{diagbox}

\makeatletter
\newtheoremstyle{mythm}{3pt}{3pt}{}{16pt}{\bfseries}{:}{.5em}{}
\renewcommand{\arraystretch}{1.2}
\theoremstyle{mythm}
\newtheorem{theorem}{Theorem}
\newtheorem{example}{Example}
\newtheorem{definition}{Definition}
\newtheorem{remark}{Remark}

\newtheorem{lemma}{Lemma}

\newcommand{\tabincell}[2]{\begin{tabular}{@{}#1@{}}#2\end{tabular}}
\begin{document}
\title{Coded Caching Scheme for Partially Connected Linear Networks Via Multi-antenna Placement Delivery Array
\author{Minquan Cheng, Yun Xie, Zhenhao Huang, Mingming Zhang, and Youlong Wu}
	\thanks{M. Cheng, Y. Xie and M. Zhang are with Guangxi Key Lab of Multi-source Information Mining $\&$ Security, Guangxi Normal University,
		Guilin 541004, China  (e-mail: chengqinshi@hotmail.com, yunxie1022@stu.gxnu.edu.cn, ztw\_07@foxmail.com,).}
	\thanks{Z. Huang and Y. Wu are with the School of Information Science and Technology, ShanghaiTech University, Shanghai 201210, China  (e-mail:  huangzhh,wuyl1@shanghaitech.edu.cn).}
}
\date{}
\maketitle
\begin{abstract}
In this paper, we study the coded caching scheme for the $(K,L,M_{\text{T}},M_{\text{U}},N)$ partially connected linear network, where there are $N$ files each of which has an equal size, $K+L-1$ transmitters and $K$ users; each user and transmitter caches at most $M_{\text{U}}$ and $M_{\text{T}}$ files respectively; each user cyclically communicates with $L$ transmitters. The goal is to design
caching and delivery schemes to reduce the transmission latency measured by the metric normalized delivery time (NDT). By delicately designing the data placement of the transmitters and users according to the topology, we show that a combinatorial structure called multiple-antenna placement delivery array (MAPDA), which was originally proposed for the multiple-input single-output broadcast channels, can be also used to design schemes for the partially connected linear network. Then, based on  existing MAPDAs and  our constructing approach, we propose new schemes that achieve the optimal NDT when $ {M_\text{T}}+ {M_\text{U}}\geq N$ and  smaller NDT than that of the existing schemes when (${M_\text{T}}+ {M_\text{U}}\leq N$, $\frac{M_\text{U}}{N}+\frac{M_\text{T}}{N}
\frac{L}{K}\left\lceil \frac{K}{L} \right\rceil \geq 1$) or  ($ {M_\text{U}}+ {M_\text{T}}< N,  \frac{K}{L}\notin\mathbb{Z}^+$). Moreover, our schemes operate in one-shot linear delivery and significantly reduce the subpacketizations compared to the existing scheme, which implies that our schemes have a wider range of applications and lower complexity of implementation.

 \end{abstract}


\section{Introduction}
The immense growth of wireless data traffic is putting incredible pressure on the wireless network, especially the high temporal variability of network traffic, resulting in congestion during peak traffic time and under utilization during off-peak time. Coded caching in \cite{MN} is an efficient solution to reduce transmission pressure by pre-populating the user's local cache with content at off-peak time and using coding theory to generate more multicast opportunities during peak traffic time.

A   caching system consists of two phases, i.e., the placement phase at off-peak traffic time and the delivery phase at peak traffic time. In the placement phase, the server places content in each user's cache without knowing future users' demands. In the delivery phase, each user requests an arbitrary
file and the server broadcasts coded packets such that each user can decode its requested file with the help of its caching contents.  The first coded caching scheme was proposed by  Maddah-Ali and Niesen in \cite{MN} for a shared-link broadcast network, where a central server containing $N$ files with equal length connects to $K$ users, each of which can cache at most $M$ files through an error-free shared link. The communication goal is to design a scheme such that the transmission cost is as small as possible.  The first coded caching scheme, referred to as the  MN scheme, achieves the minimum communication load (i.e., the number of communication bits normalized by the size of the file) within a multiplicative factor of 2\cite{YMA2019}. For the uncoded placement where each user stores directly a subset of the bits of files, the MN is exactly optimal \cite{WTP2020,YMA2018}.   Following the original caching problem,   many works have investigated the coded caching problem for a variety of network topologies, such as Device-to-Device (D2D) network~\cite{JCM,ESPA,WMDTD}, hierarchical   network~\cite{KNMD,CCHN,HHCC}, combination network~\cite{JWTCEL,CCCN,OBAC}, and arbitrary multi-server linear network~\cite{SSB},  etc. %ClThe first known coded caching scheme, referred to as the MN scheme, was proposed by Maddah-Ali and Niesen in \cite{MN}. 

%The achieved load of the MN scheme is optimal under the constraint of uncoded placement and $K\leq N$~\cite{WTP2016,WTP2020}, and generally  order optimal within a factor of $4$ \cite{GR}.  
 %When a file is requested multiple times, by removing the redundant transmissions in the MN scheme, Yu, Maddah-Ali, and Avestimehr designed a scheme that is optimal under the constraint of uncoded placement for $K>N$ \cite{YMA2018} and is order optimal within a factor of $2$ \cite{YMA2019}.  

Recently, coded caching has been widely extended to the wireless network, such as multiple-input single-output (MISO) broadcast channels \cite{EP,SCH,MB,SPSET,EBPresolving,WCC,YWCC,EPDA},  multiple-input multiple-output (MIMO) broadcast channels \cite{MIMOcacheBC'19},   single-input-single-output (SISO) interference channels \cite{NMA,HND,CAIC,FWNC,XTZ,Tao'TITic}, and  MIMO interference channels  \cite{TaoMIMOIC'17,MIMOIC'TIT23},  etc. The goal of most existing works on cache-aided wireless network is to jointly design  data placement  and physical layer delivery to improve the communication efficiency.   For instance,  \cite{NMA,HND,CAIC} proposed schemes that are (order) optimal in the sense of sum DoF for the cache-aided SISO interference channels, and \cite{Tao'TITic} established the optimal normalized delivery efficiency (NDT), a definition first introduced by \cite{NDT-Def16}, in certain cache size regions. The work in  \cite{XTZ} studied a partially connected linear network where the users can only connect with part of the transmitters, and proposed a coded caching scheme that  achieves the optimal NDT when cache memories of transmitters and users are relatively large. In addition, some works  take both communication efficiency and computational complexity into consideration. For example,  low-subpacketisation coded caching schemes generated by constructing the multiple-antenna placement delivery array (MAPDA)\footnote{The authors independently proposed the same combinatorial structure called extended placement delivery array for MISO broadcast channels.} were proposed in  \cite{MB,SPSET,YWCC,EPDA}   for MISO broadcast channels, and  one-shot linear delivery  based on interference zero-forcing was proposed in \cite{NMA} for SISO interference channels.

In this paper, we revisit the partially connected linear network  \cite{XTZ} that models a typical wireless network where some users can only communicate with a subset of transmitters due to pass loss caused by blocking objects. More specifically, we consider a wireless network with $K$ linearly aligned users, $K+L-1$ linearly aligned transmitters, and each user is locally connected to a subset of $L\in\{1,\ldots,K\}$ continuous transmitters. Let $M_{\text{T}}$ and $M_{\text{U}}$ represent the caching memory sizes of each user and transmitter, respectively. In \cite{XTZ},  a coded caching scheme based on interference alignment and interference neutralization was proposed by Xu, Tao, and Zheng, namely \emph{the XTZ scheme}, to achieve the optimal NDT when $M_{\text{T}}+M_{\text{U}}\geq N$.
Despite the optimality of the scheme in the case $M_{\text{T}}+M_{\text{R}}\geq N$, there are still several important and unresolved issues. First,  it is unknown whether the optimality still holds for the case $M_{\text{T}}+M_{\text{R}}< N$, which is a common case when each user and transmitter are equipped with insufficient size of cache memories; if the answer is not, then  how can we further improve the communication efficiency? Second, the XTZ scheme involves  high coding and computational complexity. More specifically, for the case $M_{\text{T}}/N=1/L$, the XTZ scheme splits each file into an exponentially large number of subfiles $\binom{L}{M_{\text{R}}L/N}$ and applies interference alignment, which requires each user to first wait
\begin{align*}
L \binom{L- 1}{M_{\text{R}}L/N} n^\rho+\binom{L-1}{M_{\text{R}}L/N+1}(n+1)^\rho
\end{align*}
transmission slots where $n\in\mathbb{N}$ \footnote{In general, the value $n$ should be sufficiently large such that the maximum degree of freedom is achieved.}
and $\rho=(K+L-1)(L-M_{\text{R}}L/N-1)$ to decode its desired contents. This would cause unbearable waiting latency and high computational complexity in the  transmit  beamforming vector design (See detailed discussions in Section   \ref{subsub-performance}.)

In view of the facts above, we aim to find low-complexity and communication-efficient coded caching schemes for the $(K+L-1) \times K$ partially connected wireless network, to simultaneously reduce the subpacketisation level, computational complexity, and transmission latency.  The contributions of our work can be summarized as follows.

$\bullet$ We first prove that MAPDA can also be used to design coded caching schemes for the partially connected wireless network, and then propose  new coded caching schemes based on   existing MAPDAs, which are listed in Table \ref{tab-Theorem-2}. Note that the MAPDA was originally proposed for wireless network where all users connect with all transmitters, and directly applying MAPDA to the partially connected wireless network would lead to the users missing some signals due to  disconnecting links. To address this issue, we delicately design the data placement at the transmitters via a cyclic-based MAPDA such that any $L$ consecutive transmitters can store and deliver all required contents to their connected users, and then globally design all the users's placement based on an integral  MAPDA. This enables our scheme to simultaneously deliver all desired files to all users and achieve larger multicast opportunities compared to the XTZ scheme.

$\bullet$  Compared with the XTZ scheme, our schemes achieve smaller NDTs when 1) ${M_\text{T}}+ {M_\text{U}}\leq N$, $\frac{M_\text{U}}{N}+\frac{M_\text{T}}{N}
\frac{L}{K}\left\lceil \frac{K}{L} \right\rceil \geq 1$ or 2)  $ {M_\text{U}}+ {M_\text{T}}< N,   {K}/{L}\notin\mathbb{Z}^+$;   the same  NDT as the XTZ scheme when $ {M_\text{U}} + {M_\text{T}} \geq N$ (optimal NDT is achieved) or $ {M_\text{U}}+ {M_\text{T}}< N,   {K}/{L}\in\mathbb{Z}^+$;   and slightly larger NDTs than the XTZ scheme when  $ {M_\text{U}}+ {M_\text{T}}< N$ and $ L{M_\text{T}}=N$ but preserving much lower decoding complexity due to the one-shot delivery strategy (see Table \ref{tab-relationship}).

$\bullet$  Unlike the XTZ scheme where the subpacketization  grows exponentially with $L$, our schemes significantly reduce the subpacketization that linearly increases with $K$. Moreover, our scheme enables independent data placement between the transmitter and user, while   the XTZ scheme requires data placements among all nodes to be dependent on each other. Finally,  our schemes operate in a one-shot delivery strategy for all cases, while the XTZ scheme needs all users to wait for long transmission slots and then decode the information for some cases (e.g., $LM_T/N=1$).  These facts indicate that our schemes have a wider range of applications and lower complexity of implementation (see Section \ref{subsub-delivery}).

	\paragraph*{Paper Organization}	
	The rest of this paper is organized as follows. Section~\ref{sec-pre} describes the system model. Section \ref{sec-MAPDA-PC} present MAPDA for the partially connected linear network. Some proofs can be found in Section \ref{sec-proof-main-result} and Appendices.

{\bf Notations:} In this paper, the following notations will be used unless otherwise stated.

$\bullet$ $[a:b]:=\left\{ a,a+1,\ldots,b\right\}$ and $[a]:=\left\{ 1,2,\ldots,a\right\}$.  $|\cdot|$ denotes the cardinality of a set.

$\bullet$  We use the notation $a\mid q$ if $a$ is divisible by $q$ and $a\nmid q$ otherwise. If $a$ is not divisible by $q$, $\langle a\rangle _q$ denotes the least non-negative residue of $a$ modulo $q$; otherwise, $\langle a\rangle _q:=q$.

$\bullet$ $gcd(a,b)$ denotes the greatest common divisor of $a$ and $b$.

$\bullet$  Let $\mathcal{B}=\{b_1,b_2,\ldots,b_{n}\}$ be a  set with $b_1<b_2<\ldots<b_n$, for any $i\in[n]$, $\mathcal{B}[i]$ denotes the $i^{\text{th}}$ smallest element of $\mathcal{B}$, i.e., $\mathcal{B}[i]=b_i$.

$\bullet$  For any positive integers $n$ and $t$ with $t<n$, let ${[n]\choose t}=\{\mathcal{T}\ |\   \mathcal{T}\subseteq [n], |\mathcal{T}|=t\}$, i.e., ${[n]\choose t}$ is the collection of all $t$-sized subsets of $[n]$.

$\bullet$ Let ${\bf a}$ be a vector with length $n$, for any $i\in[n]$, ${\bf a}[i]$ denotes the $i^{\text{th}}$ coordinate of ${\bf a}$.
For any subset $\mathcal{T}\subseteq [n]$, ${\bf a}[\mathcal{T}]$ denotes a vector with length $|\mathcal{T}|$ obtained by taking only the coordinates with subscript $i\in \mathcal{T}$.

$\bullet$ Given any $F\times m$ array $\mathbf{P}$, for any integers $i\in[F]$ and $j\in [m]$, $\mathbf{P}(i,j)$ represents the element located in the $i^{\text{th}}$ row and the $j^{\text{th}}$ column of $\mathbf{P}$; $\mathbf{P}(\mathcal{ V},\mathcal{T})$ represents the subarray generated by the row indices in $\mathcal{V}\subseteq [F]$ and the columns indices in $\mathcal{T}\subseteq [m]$. In particular let $\mathbf{P}([F],\mathcal{T})$  be shortened by $\mathbf{P}(\cdot,\mathcal{T})$ and  $\mathbf{P}(\mathcal{V},[m])$  be shortened by $\mathbf{P}(\mathcal{V},\cdot)$.

\section{ Partially Connected Networks placement Delivery Array}
\label{sec-pre}

\subsection{System Model}
\label{PCPDA}
\begin{figure}
\centering
		\includegraphics[scale=0.3]{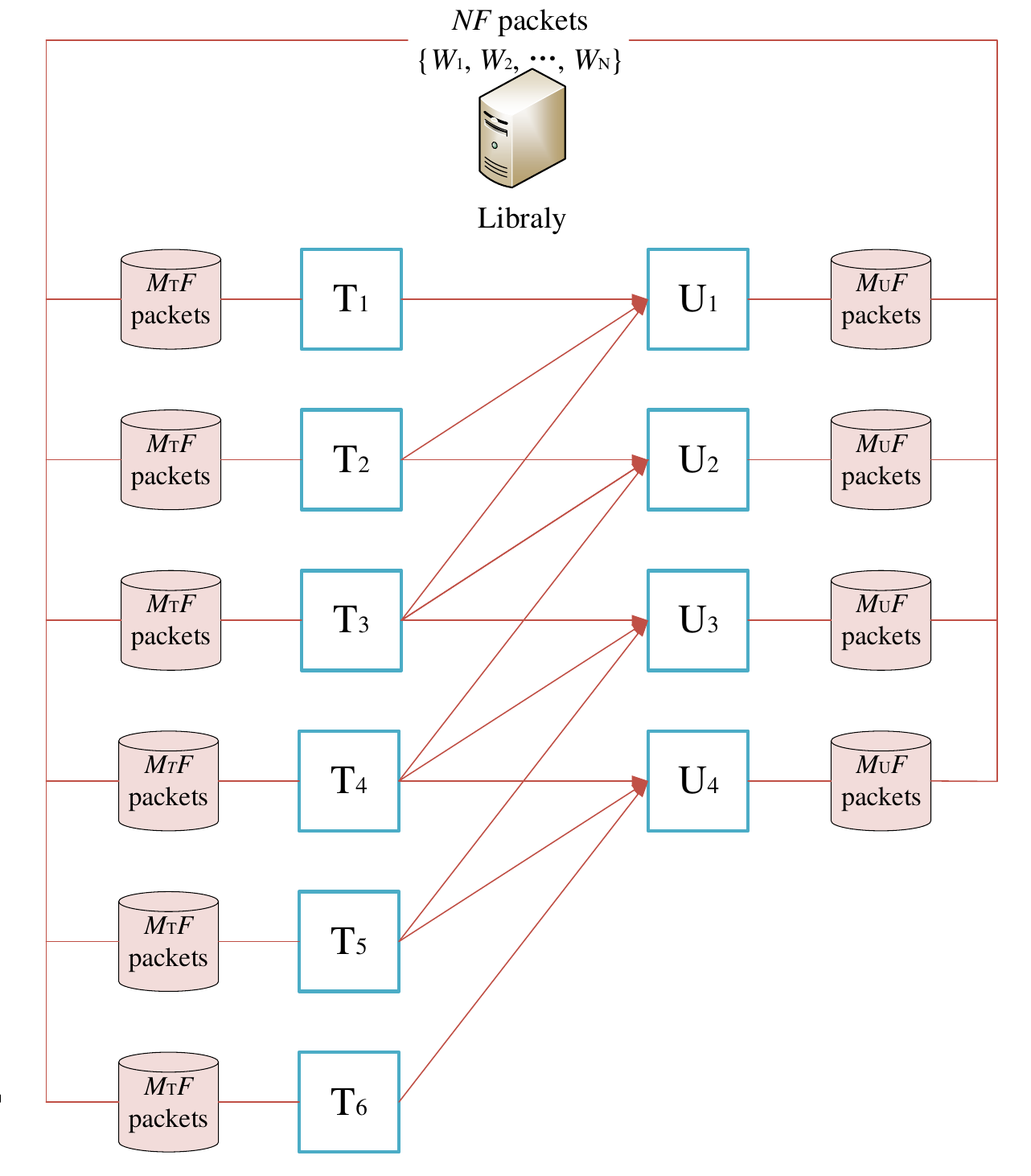}
		\caption{$6\times 4$ partially connected linear network with user connectivity $L=3$.}
\label{fig-model2}
\end{figure}

Consider a $(K+L-1)\times K$ partially connected linear network (see Fig. \ref{fig-model2}), where there is a library of $N$ files $\mathcal{W}=\{W_1,\ldots,W_N\}$ each of $V$-bit length, $K+L-1$ linearly aligned transmitters  denoted by T$_1$, T$_2$, $\ldots$, T$_{K+L-1}$, $K$ linearly aligned users denoted by U$_1$, U$_2$, $\ldots$, U$_K$, and each user U$_k$ is connected to $L$ consecutive transmitters T$_k$, T$_{k+1}$, $\ldots$, T$_{k+L-1}$ where $k\in[K]$ and $L\leq K$. Here, $L$ is referred to as \emph{user connectivity}. Fig. \ref{fig-model2} shows an example of the linear network with  $K = 4$ and $L = 3$ where each transmitter is equipped with a cache of finite size and has a single antenna.  A $(K,L,M_{\text{T}},M_{\text{U}},N)$ coded caching scheme contains two phases.
\subsubsection{Placement phase} In this paper, we consider the uncoded placement where every node directly caches a subset of the library bits. Each file is divided into $F$ packets, i.e., $W_{n}=(W_{n,1},W_{n,2},\ldots,W_{n,F})$, where each packet $W_{n,f}\in\mathbb{F}_{2^B}$ for $n\in [N], f\in[F]$. Here $B$ represents the size of each packet. Clearly, we have $V=FB$. Each transmitter and each user caches some packets of $\mathcal{W}$ with a size of at most $M_{\text{T}}F$ packets and $M_{\text{U}}F$ packets, respectively. Denote the cached contents at transmitter T$_j$ where $j\in[K+L-1]$ and user  U$_k$ where $k\in[K]$ as $\mathcal{Z}_{\text{T}_j}$ and  $\mathcal{Z}_{\text{U}_k}$, respectively.%{\color{blue}This placement is called uncoded placement.  }

We assume that the placement is performed without knowing users' later demands.

\subsubsection{Delivery phase} Each user U$_k$ requests for an arbitrary  file $W_{d_k}$ from the library $d_k\in[N]$, for $k\in[K]$.   Let $\mathbf{d}\triangleq (d_1,d_2,\ldots,d_K)$ denote the demand vector. According to the users' demands and caches, the server transmits coded packets through $L$ antennas. More precisely, the server first uses a code for the Gaussian channel with the rate
\begin{IEEEeqnarray}{rCl}\label{eq-rate-packet}
\frac{B}{\tilde{B}}=\log P+o(\log P)%(bit per complex symbol)
\end{IEEEeqnarray} to encode each packet  to a \emph{coded packet} as  $\tilde{ {W}}_{n,f}=\psi( {W}_{n,f})\in\mathbb{C}^{\tilde{B}}$, where $\psi$ is the coding scheme for the Gaussian channel, e.g., random Gaussian coding. Here each coded packet contains $\tilde{B}$ complex symbols and carries one degree-of-freedom (DoF).	
	 The whole communication process contains $S$ blocks, each of which consists of $\tilde{B}$  complex symbols (i.e.,  $\tilde{B}$ time slots). In each block $s\in [S]$, the communication goal is to deliver a subset of the requested packets, denoted by $\mathcal{D}_s = \{\tilde{W}_{d_{k_1},f_1},\ldots,\tilde{W}_{d_{k_{|\mathcal{D}_s |}},f_{|\mathcal{D}_s |}}\}$, to a subset of users $\mathcal{K}_s = \{k_1,\ldots, k_{|\mathcal{D}_s |}\}$. Assume that the user U$_{k_i}$ requests the packet $\tilde{W}_{d_{k_i},f_i}$ for each $i\in |\mathcal{D}_s|$. In this paper we only consider \emph{linear}  coding schemes in the delivery phase. In  each block $s\in[S]$, each transmitter T$_j$ where $j\in [K+L-1]$  sends $\mathbf{x}^{(s)}_{j}\in \mathbb{C}^{\tilde{B}}$, which is linear combinations of the coded plackets, i.e.,
\begin{align*}
\mathbf{x}^{(s)}_{j} = \sum_{i\in[|\mathcal{D}_s|]} v^{(s)}_{j,k_i}  \tilde{W}_{d_{k_i},f_i},
\end{align*}
where $v^{(s)}_{j,k_i}$ is the complex beam forming coefficient and can be any complex value if the packet $W_{d_{k_i},f_i}$ is cached by transmitter T$_j$, otherwise $v^{(s)}_{j,k_i}=0$ for each $i\in [|\mathcal{D}_s|]$. Then each user $\text{U}_k$, $k\in \mathcal{K}_{s}$ can receive the following signal
\begin{align}
\label{eq-signal-partially}
\mathbf{y}^{(s)}_{k}=\sum_{j=k}^{k+L-1}h^{(s)}_{k,j}\mathbf{x}^{(s)}_{j}+
\mbox{\boldmath$\epsilon$}^{(s)}_k,
\end{align}through the interference channel
where $h^{(s)}_{k,j}\in \mathbb{C}$ denotes the channel coefficient from transmitter T$_j$ to user U$_k$, which is independent and identically distributed in $\mathbb{C}$.
User $\text{U}_k\in\mathcal{K}_s$ can decode the following coded signal
\begin{align*}
\tilde{W}_{d_k,f} +\mbox{\boldmath$\epsilon$}^{(s)}_k.
\end{align*}based on its local caches and received signal $\mathbf{y}^{(s)}_{k}$. By assuming  $P$ is large enough, the coded packet $\tilde{W}_{d_k,f}$ can be decoded with an error probability exponentially decreasing to zero. To evaluate the transmission efficiency of the scheme, we adopt the same metric \emph{normalized delivery latency} (NDT) as in \cite{FWNC,XTZ}, which is defined as
\begin{eqnarray}\label{eq-original-NDT}
\tau(M_{\text{T}},M_{\text{U}})\triangleq
\lim_{P\to\infty}\lim_{V\to\infty}\sup\frac{\max_{{\bf d}\in[N]^K}T}{V/\log P},
\end{eqnarray} where $T$ is the total time slots in the whole communication process. Since each file contains $F$ packets, each of which has $B$ bits and there are in total $S\tilde{B}$ time slots, \eqref{eq-original-NDT} can be written as
\begin{align}\label{eq-comput-NDT}
\tau&= \lim_{P\to\infty}\lim_{V\to\infty}\frac{S\tilde{B}}{BF/\log P}\nonumber\\
&=\lim_{P\to\infty}\frac{S}{F}\cdot\frac{\log P}{\log P+o(\log P)}
=\frac{S}{F}.
\end{align}
From \eqref{eq-comput-NDT}, NDT can represent the maximal normalized number of transmitted files over all possible demands in the interference channel and the high signal-to-noise ratio (SNR) regime. We prefer to design a scheme with the optimal NDT defined as
\begin{align*}
\tau^*(M_{\text{T}},M_{\text{U}})\triangleq\inf\{
\tau(M_{\text{T}},M_{\text{U}})\ |\
\tau(M_{\text{T}},M_{\text{U}})\ \text{is achievable}\}.
\end{align*}
\begin{remark}\label{remark-Def-DoF}In \cite{NMA}, the authors applied the metric sum DoF to measure the communication efficiency, which is defined as the total transmitted requested packet bits per time slot normalized by $\log P$, i.e.,
\begin{align}
\label{eq-def-DoF}
\text{Sum-DoF} = \lim_{P\to\infty}\frac{K(1-M_\text{U}/N)BF}{S\tilde{B}\log P}
 = \frac{K(1-M_\text{U}/N)}{\tau},
\end{align}where the last equality holds by \eqref{eq-rate-packet} and  \eqref{eq-comput-NDT}.
\end{remark}

The first coded caching scheme for the partially connected linear network was proposed in \cite{XTZ}, where the following result was given.

\begin{lemma}[NDT of the XTZ scheme \cite{XTZ}]
\label{le-Tao}
For the cache-aided $(K+L-1)\times K$ partially connected linear network, there exists a $(K, L, M_{\text{T}},M_{\text{U}},N)$ coded caching scheme with the NDT as follows.
\begin{align}
\label{eq-Tao-ndt}
&\tau_{\text{XTZ}}(M_{\text{T}},M_{\text{U}})\nonumber\\
=&\left\{
\begin{array}{cc}
(1-\frac{1}{L}+\frac{1}{ \frac{M_{\text{U}}L}{N} + 1 })(1-\frac{M_{\text{U}}}{N}) & \text{\tabincell{c}{if $\frac{ M_{\text{T}}L }{ N }=1$,\\
$\frac{ M_{\text{U}}L }{ N }\in[0:L-1]$,}}\ \ \ \ \ \\[0.4cm]
\frac{ 1-\frac{M_{\text{U}}}{N}}{ \min\{\frac{M_{\text{T}}}{N}+\frac{M_{\text{U}}}{N}, 1 \} }\ \ \ \ \ \ \ \ \ \ \ \ \ \ \ & \text{\tabincell{c}{if   $\frac{M_{\text{T}}L}{N}\in[2:L]$, \\ $\frac{M_{\text{U}}L}{N}\in[0:L-1]$.}}
		\end{array}
		\right.
	\end{align}
\end{lemma}
It is worth mentioning that the authors in \cite{XTZ} showed that when $\frac{M_\text{T}}{N}+\frac{M_\text{T}}{N} \geq 1$, the scheme in Lemma \ref{le-Tao} achieves the optimal NDT.
In this paper, we aim to propose communication-efficient and low-complexity schemes that improve the scheme in Lemma  \ref{le-Tao} for the case $\frac{M_\text{T}}{N}+\frac{M_\text{T}}{N} < 1$.

\subsection{Multi-antenna Placement Delivery Array}
The authors in \cite{YWCC} proposed multiple-antenna placement delivery array  (MAPDA) to characterize the placement strategy and delivery strategy for the MISO caching system. In this section, we will introduce MAPDA that will be helpful in generating schemes for the partially connected linear network.
\begin{definition}[\cite{YWCC}]\rm
		\label{def-MAPDA}
		For any positive integers $r$, $K$, $F$, $Z$ and $S$, an $F\times K$ array $\mathbf{Q}$ that is composed of $``*"$ and $[S]$ is called $(r,K,F,Z,S)$ multiple-antenna placement delivery array (MAPDA) if it satisfies the following conditions
		\begin{itemize}
\item [C$1$.] The symbol $``*"$ appears $Z$ times in each column;
\item [C$2$.] Each integer occurs at least once in the array;
			\item[C$3$.] Each integer $s$ appears at most once in each column;
			\item[C$4$.] For any integer $s\in[S]$, define  $\mathbf{Q}^{(s)} $
			to be the subarray of $\mathbf{Q}$ including the rows and columns containing $s$, and let $r'_s\times r_s$ denote the dimensions of $\mathbf{Q}^{(s)}$.  The number of integer entries in each row  of $\mathbf{Q}^{(s)}$ is less than or equal to $r$, i.e.,
			\begin{eqnarray}\label{C4}
				\left|\{k_1\in [r_s]  |\ \mathbf{Q}^{(s)}(f_1,k_1)\in[S]\}\right|\leq r, \ \forall f_1 \in [r'_s].
			\end{eqnarray}
		\end{itemize}
		\hfill $\square$
	\end{definition}
If each integer appears $g$ times in the $\mathbf{Q}$, then $\mathbf{Q}$ is a g-regular MAPDA, denoted by  $g$-$(r,K,F,Z,S)$ MAPDA.

\begin{example}\label{ex-1}
We can check that the $5\times 5$ array $\mathbf{Q}$ is a $g$-$(r,K,F,Z,S)$ = $5$-$(4,5,5,1,4)$ MAPDA.
\begin{align*}%\label{ex-Or-PCPDA}
\mathbf{Q}=\left(\begin{array}{ccccc}
				* & 1 & 1 & 1&1 \\[-0.15cm]
				1 & * & 2 & 2&2 \\[-0.15cm]
				2 & 2 & * & 3&3 \\[-0.15cm]
				3 & 3 & 3 & *&4\\[-0.15cm]
                4 & 4 & 4 & 4&*
			\end{array}\right).
\end{align*} For instance, when $s =1$, we have the following sub-array.
\begin{align*}
\mathbf{Q}^{(1)}=\left(\begin{array}{ccccc}
				* & 1 & 1 & 1&1 \\
				1 & * & 2 & 2 &2
\end{array}\right).
\end{align*} It can be seen that each row of $\mathbf{Q}^{(1)}$ contains $4$ integer entries and no more than $r=4$ integer entries. Hence, $\mathbf{Q}^{(1)}$ satisfies the condition C$4$ of Definition \ref{def-MAPDA}.
\hfill $\square$
\end{example}

In a $(r,K,M,N)$ MISO caching system, there is a server containing  $N$ files with the same size, and $K$ users each of which has a cache with the capacity of $M$ files through $r$ antennas over the interference channel. Given a $(r,K,F,Z,S)$ MAPDA $\mathbf{Q}$, we can obtain a $(r,K,M,N)$  scheme for MISO caching system in the following two phases.

$\bullet$  {\bf Placement phase:} The $K$ columns and $F$ rows denote the users and packets of each file, respectively. Specifically, the server divides each file into $F$ packets; the entry $\mathbf{Q}(f,k)=*$ means that the $f^{\text{th}}$ packets of all files are cached by user $k$. Each user caches $M=\frac{ZN}{F}$ files by Condition C$1$ of the Definition \ref{def-MAPDA}. So, if $\mathbf{Q}(f,k)$, for $f\in[F]$ and $k\in[K]$, is an integer $s\in[S]$, then the $f^{\text{th}}$ packet of each file is not stored by user $k$.
    Clearly, the placement strategy of the scheme realized by MAPDA is called uncoded cache placement.

$\bullet$  {\bf Delivery phase:} The integer $s\in[S]$ represents delivery strategy at the block $s$. For any demand vector ${\bf d}$, at block $s$, the server first chooses a pre-coding matrix for $r$ antennas to encode the requested packets indicated by $s$, and then sends coded packets to the users.

{\begin{table*}[http] 	
\caption{Existing $(r,K,F,Z,S)$ MAPDAs with $K\in \mathbb{N}^+$, the sum-DoF $\min\{r+ \frac{KM}{N},K\}$, $r\in \mathbb{N}^+$ antennas and memory ration $\frac{M}{N}=\frac{z}{K}$.}
\centering
\begin{spacing}{1.5}
\begin{tabular}{|c|c|c|c|c|c|c|}	\hline
Index&MAPDAs &   F &  $Z$ &  $S$\\	\hline
		
1&\cite{NMA}: $z+r \leq K$&$\binom{K}{z}\frac{ z!(K-z-1)!}{(K-z-r)!}$&$\binom{K}{z}\frac{ z\cdot z!(K-z-1)!}{K(K-z-r)!}$&$\frac{(K-z)z!(K-z-1)!}{(r+ z)(K-z-r)!}\binom{K}{z}$ \\		\hline
		
2&\cite{EP}: $\frac{K}{r},\frac{z}{r}\in\mathbb Z^+$& $\binom{K/r}{z/r}$&$\frac{z}{K}\binom{K/r}{z/r}$&$\binom{K/r}{z/r+1}$\\	\hline		

3&\cite{SCH}: $z+r \leq K$ &$\binom{K}{z}\binom{K-z-1}{r-1}$&$\frac{z}{K}\binom{K}{z}\binom{K-z-1}{r-1}$&$\frac{(K-z)\binom{K}{z}\binom{K-z-1}{r-1}}{r+ z}$\\ \hline
			
4&\cite{MB}: $\frac{z+r}{z+1}\in\mathbb{Z}^+$&$\binom{K}{z}$&$\frac{z}{K}\binom{K}{z}$ &$\frac{z+1}{(z+r)}\binom{K}{z+1}$\\	\hline

5&\tabincell{c}{\cite{YWCC}:$z+r<K$, $m\leq r$,\\[-0.3cm]  $l=\frac{m}{\gcd(m,r-m)}$, $\beta =$\\[-0.3cm] $(\text{sgn}(\frac{z}{m}+1,\frac{m}{r})+\frac{r-m}{m})l$} &$\beta {\frac{K}{m}\choose \frac{z}{m}}$& $\frac{\beta z}{K} {\frac{K}{m}\choose \frac{z}{m}}$ & \tabincell{c}{$ \text{sgn}(\frac{z}{m}+1,\frac{m}{r})\cdot$\\ $l\cdot\frac{K-z}{z+m} {\frac{K}{m}\choose \frac{z}{m}}$}\\	
\hline

6&\tabincell{c}{\cite{EPDA}: $z+r\leq K$,\\[-0.3cm]
 $\alpha=\gcd(K,z,r)$} &$\frac{z+r}{\alpha} \binom{\frac{K}{\alpha}}{\frac{z+r}{\alpha}}$& $\frac{z}{\alpha}\binom{\frac{K}{\alpha}-1}{\frac{z+r}{\alpha}-1}$&
$\frac{K-z}{\alpha}\binom{\frac{K}{\alpha}}{\frac{z+r}{\alpha}}$\\	
\hline

7&\cite{YWCC}: $r=K-z$&$K$& $z$&$K-z$\\	 \hline

8&\tabincell{c}{\cite{SPSET}: $z\leq r$\\[-0.3cm] $\alpha=\gcd(K,z,r)$} &$\frac{K(z+r)}{\alpha^2}$& $\frac{z(z+r)}{\alpha^2}$&$\frac{K(K-z)}{\alpha^2}$\\	\hline
9&\tabincell{c}{\cite{EPDA}:$K=nz+(n-1)r$ \\[-0.3cm]$n\geq 2$, $r\geq z$,\\[-0.3cm] $\alpha=\gcd(K,z,r)$} &$\frac{K}{\alpha}$& $\frac{z}{\alpha}$& $\frac{K-z}{\alpha}$\\	 \hline
10&\tabincell{c}{\cite{EPDA}:  $r=K-z$,\\[-0.3cm] $\alpha=\gcd(K,z,r)$} &$\frac{K}{\alpha}$& $\frac{z}{\alpha}$& $\frac{K-z}{\alpha}$\\	 \hline
\multicolumn{4}{l}{\small * $\text{sgn}(x,y)$ equals to $1$ if $ y=1$, and $x$ otherwise. }\\
		\end{tabular}
	\end{spacing}	
\label{tab-compare-1}
\end{table*}
}
In \cite{YWCC}, the authors pointed out that 1) at each block the server multicasts $r_s$  packets requested by $r_s$ different users by Conditions C2-3 of Definition \ref{def-MAPDA}; 2) the condition C$4$ of Definition \ref{def-MAPDA} ensures that at each block $s\in [S]$, the server can always find the pre-coding matrices such that each user can recover its requested packet. So, the delivery strategy of the scheme realized by MAPDA is a one-shot linear delivery. Then we can obtain the following result.
\begin{lemma}[\cite{YWCC}]\label{lemma-MAPDA-CCS}
Given $(r,K,F,Z,S)$ MAPDA $\mathbf{Q} $, there exists an  $F$-division scheme for the $(r,K,M,N)$ multiple antennas coded caching  problem with memory ratio $\frac{M}{N}=\frac{Z}{F}$, sum-DoF $\frac{K(F-Z)}{S}$ and subpacketization $F$.
\hfill $\square$
\end{lemma}

Under the constraints of uncoded cache placement and one-shot linear delivery, the maximum sum-DoF is upper bounded by $\min\{K,\frac{KM}{N}+r\}$~\cite{EBPresolving}, which is also an upper bound on the sum-DoF achieved by the caching schemes from MAPDA.
\begin{lemma}[\cite{EBPresolving,YWCC,EPDA}]\label{le-MAPDA-CCS}
For the $(r,K,M,N)$ multiple-antenna coded caching scheme with memory ratio $\frac{M}{N}=\frac{Z}{F}$ generated by a $(r,K,F,Z,S)$ MAPDA, the sum-DoF is no more than $\min\{\frac{KZ}{F}+r=\frac{KM}{N}+r,K\}$.
		\hfill $\square$
\end{lemma}

In the literature, the schemes in~\cite{NMA,EP,SPSET,MB,STSK} can be represented by MAPDAs. So we summarize all the schemes which achieve the  sum-DoF $\min\{r+ \frac{KM}{N},K\}$ by MAPDAs in Table~\ref{tab-compare-1}.

In Table \ref{tab-compare-1}, it is not difficult to check that the subpacketizations $F$ of the first six MAPDAs are exponent with the number of users; the subpacketizations of the last five MAPDAs are linear with the number of users; the second MAPDA is a special case of the fifth scheme; the subpacketizations of the third and fourth MAPDAs are larger than that of the fifth scheme. So in the following we will only use the fifth, sixth, seventh and eighth MAPDAs in Table \ref{tab-compare-1}.

The authors in \cite{YWCC} also pointed out that MAPDA can also be used for the SISO cache-aided interference channels by viewing the transmitters as transmit antennas.  This is because  the channel coefficient between any transmitter and user can be chosen independently and identically distributed (i.i.d) from $\mathbb{C}$, which enables us to always find a pre-coding matrix for all the transmitters. However, the MAPDA can not be directly applied to design coding schemes for the partially connected linear network since channel coefficients $h^{(s)}_{k,j}$  can be chosen independently and identically from $\mathbb{C}$ only if the transmitter T$_{j}$ connects to the user U$_{k}$, and $h^{(s)}_{k,j}=0$ otherwise. This means that directly applying MAPDA will lead the users not to receive the signals carrying their desired packets, due to the disconnecting link $h^{(s)}_{k,j}=0$.

\section{MAPDA for Partially Connected Linear Interference Networks}
\label{sec-MAPDA-PC}
In this section, we will show that MAPDA can also be used to generate a coded caching scheme for the partially connected linear network by our novel construction and the Schwartz-Zippel Lemma \cite{DL}, i.e., the following result which is proved in Section \ref{sec-proof-main-result}.

\begin{theorem}\label{th-PCPDA-CCS}
Given a $(r,K,F_1,Z_1,S_1)$ MAPDA, there exists a $(K,L,M_{\text{T}},M_{\text{U}},N)$ coded caching scheme for the $(K+L-1)\times K$ partially connected linear network achieving the NDT $\tau = \frac{S_1}{F_1}$
with $\frac{L M_{\text{T}}}{N}=\frac{r}{\lceil K/L\rceil}\in\mathbb{Z}^+$, $\frac{M_{\text{U}}}{N}=\frac{Z_1}{F_1}$, and subpacketization $F=LF_1$.
\hfill $\square$
\end{theorem}

By Theorem \ref{th-PCPDA-CCS} and the fifth, sixth, seventh and eighth MAPDAs in Table \ref{tab-compare-1}, we have the following schemes for the $(K+L-1)\times K$ partially connected linear network.
\begin{theorem}
\label{PC-mapda}
For any positive integers $K$, $L$, $m$ and $N$, there exist four $(K,L,M_{\text{T}},M_{\text{U}},N)$ coded caching schemes for the $(K+L-1)\times K$ partially connected linear network with $\frac{M_{\text{T}}}{N}\in \{\frac{1}{L},\ldots,\frac{L-1}{L},1\}$, $\frac{M_{\text{U}}}{N}\in \{0,\frac{1}{K}, \ldots,\frac{K-1}{K},1\}$, NDT $\tau_{\text{new}}=\frac{K(1-M_{\text{U}}/N)}{\min\{K,\frac{KM_{\text{U}}+M_{\text{T}}L\lceil K/L\rceil}{N}\}}$, subpacketization and their parameter limitations listed in Table \ref{tab-Theorem-2}.

{\begin{table*}[http!]
	\caption{The new schemes with subpacketization and parameter limitations in Theorem \ref{PC-mapda} where $m\in \mathbb{Z}^{+}$ and $m|K$.}
	\centering
	\label{tab-Theorem-2}
			\begin{tabular}{|c|c|c|c|}
				\hline
				New schemes&Subpacketization & Original MAPDA&Parameter limitations\\
				\hline
Scheme 1 & $L K\cdot\min\{K,\frac{KM_{\text{U}}+M_{\text{T}}L\lceil K/L\rceil}{N}\}/\alpha^2$&\cite{SPSET} & $M_{\text{T}}\cdot \frac{L}{K} \lceil \frac{K}{L} \rceil \geq M_{\text{U}}$\\ \hline
						
Scheme 2&$LK$ & \tabincell{c}{
\multirow{2}{*}{\cite{YWCC}}}& \tabincell{c}{$\frac{M_{\text{U}}}{N} + \frac{M_{\text{T}}}{N}\cdot\frac{L}{K}\lceil\frac{K}{L}\rceil \geq 1$}\\ \cline{1-2}\cline{4-4}

Scheme 3& $L\beta\binom{K/m}{KM_{\text{U}}/(mN)}$ &&$\frac{M_{\text{U}}}{N} + \frac{M_{\text{T}}}{N}\cdot\frac{L}{K}\lceil\frac{K}{L}\rceil < 1$ \\ \hline

Scheme 4&$L\frac{KM_{\text{U}}+M_{\text{T}}L\lceil K/L\rceil}{\alpha N} \binom{K/\alpha}{\frac{KM_{\text{U}}+M_{\text{T}}L\lceil K/L\rceil}{\alpha N}}$&\cite{EPDA} &$\frac{M_{\text{U}}}{N} + \frac{M_{\text{T}}}{N}\cdot\frac{L}{K}\lceil\frac{K}{L}\rceil < 1$ \\ \hline
\multicolumn{4}{l}{\small * $\text{sgn}(x,y)$ equals to $1$ if $ y=1$, and $x$ otherwise. }\\
\multicolumn{4}{l}{\small *
$\alpha=\gcd\left(K,\frac{KM_{\text{U}}}{N},\frac{M_{\text{T}}L\lceil K/L\rceil}{N}\right)$, $\beta =\frac{m}{\gcd(m,\frac{M_{\text{T}}L\lceil K/L\rceil}{N}-m)}\cdot\left(
\text{sgn}\left(\frac{KM_{\text{U}}}{mN}+1,
\frac{mN}{LM_{\text{T}}\lceil K/L\rceil}\right)
+\frac{M_{\text{T}}L\lceil K/L\rceil}{mN}-1\right)$.}
			\end{tabular}
\end{table*}
}
\end{theorem}

\subsection{Performance Analyses}
\label{subsub-performance}
In this subsection, we will show the advantages of our schemes compared to the XTZ scheme in \cite{XTZ} from the point of view of subpacketization, NDT, and complexity in the delivery phase, respectively. Prior to the comparisons, we first introduce the data placement  and the delivery strategies of the scheme in \cite{XTZ}.
{
\begin{table*}[http!]
	\centering
	\caption{The NDT $\tau_{\text{new}}$ of our scheme in Theorem \ref{PC-mapda} and the NDT $\tau_{\text{XTZ}}$ of the XTZ scheme where $L>1$.}\label{tab-relationship}
	\begin{tabular}{|c|c|c|c|c|}
		\hline
		\multicolumn{4}{|c|}{Parameter limitations }&NDTs\\ \hline
		$\frac{M_\text{U}}{N}+\frac{M_\text{T}}{N}\geq 1$&\multicolumn{3}{c|}{ }& $\tau_{\text{new}} = \tau_{\text{XTZ}}$\\ \hline
		
		&$\frac{M_\text{U}}{N}+\frac{M_\text{T}}{N}
		\frac{L}{K}\left\lceil \frac{K}{L} \right\rceil \geq 1$&\multicolumn{2}{c|}{ }&
		\\ \cline{2-4}
		
		& & \multirow{2}{*}{$\frac{LM_\text{T}}{N} \in[2:L-1]$} & $L\nmid K$ & $\tau_{\text{new}} < \tau_{\text{XTZ}}$\\ \cline{4-5}

		$\frac{M_\text{U}}{N}+\frac{M_\text{T}}{N}<1$&$\frac{M_\text{U}}{N}+\frac{M_\text{T}}{N}
		\frac{L}{K}\left\lceil \frac{K}{L} \right\rceil < 1$  &&$L\mid K$& $\tau_{\text{new}} = \tau_{\text{XTZ}}$\\ \cline{3-5}
		
		& & $LM_\text{T}=N$&  & $\tau_{\text{new}} >  \tau_{\text{XTZ}}$\\ \hline
	\end{tabular}
\end{table*}
}
\subsubsection{The placement strategy for users of XTZ scheme in \cite{XTZ}}
\label{subsub-placement-users} Given the data placement in   $(L,M_{\text{U}},N)$ MN scheme (i.e., the scheme for the shared-link network with $L$ users each with a cache size of $M_{\text{U}}$ files), each user in a partially connected network consecutively chooses a placement method such that the  placement method chosen for any $L$ consecutive of users follows exactly the placement strategy of the $(L,M_{\text{U}},N)$ MN scheme. Here we summarize the subpacketization of XTZ scheme in \cite{XTZ} as follows.
\begin{align}\label{eq-subpacket-XTZ}
	\begin{cases}
		\binom{L}{\frac{M_{\text{U}} L}{N}}L, &\!\! \frac{LM_\text{T}}{N}=1\\
		\binom{L}{\frac{M_\text{U}L}{N}}(L-\frac{M_\text{U}L}{N}), &\!\! \frac{LM_\text{T}}{N}\in[2:L], \frac{M_\text{T}}{N}+\frac{M_\text{U}}{N}\geq 1,\\
		\binom{L}{\frac{M_\text{U}L}{N}}\!\binom{L-\frac{M_\text{U}L}{N}-1}{L\frac{M_\text{T}L}{N}-1}\!(L\!-\frac{M_\text{U}L}{N}\!), &\!\! \frac{LM_\text{T}}{N}\in[2:L], \frac{M_\text{T}}{N}+\frac{M_\text{U}}{N}< 1.
	\end{cases}
\end{align}From \eqref{eq-subpacket-XTZ}, we observe that the XTZ scheme requires a subpacketization that grows exponentially with $L$.

By Table \ref{tab-Theorem-2}, the subpacketization of our schemes $F$ is small or linear with the number of users $K$ for some parameters, instead of exponentially increasing in the XTZ scheme, which demonstrates the advantage of the small subpacketizations.

Unlike the XTZ scheme that takes  $L$ consecutive users at a time and uses $L$-user MN data placement, our data placement is determined by the stars in a $(r,K,F_1,Z_1,S_1)$ MAPDA, which implies that our schemes globally design the placement and delivery among all the $K$ users. Naturally, our schemes could improve the NDT of the XTZ scheme for some parameters as the global design among all users' data placement could potentially create larger multicast opportunities. By theoretical comparisons, we have Table \ref{tab-relationship} whose proof is included in Appendix \ref{appendix-coro-NDT}.

\begin{figure}[http!]
	\centering
	\includegraphics[scale=0.25]{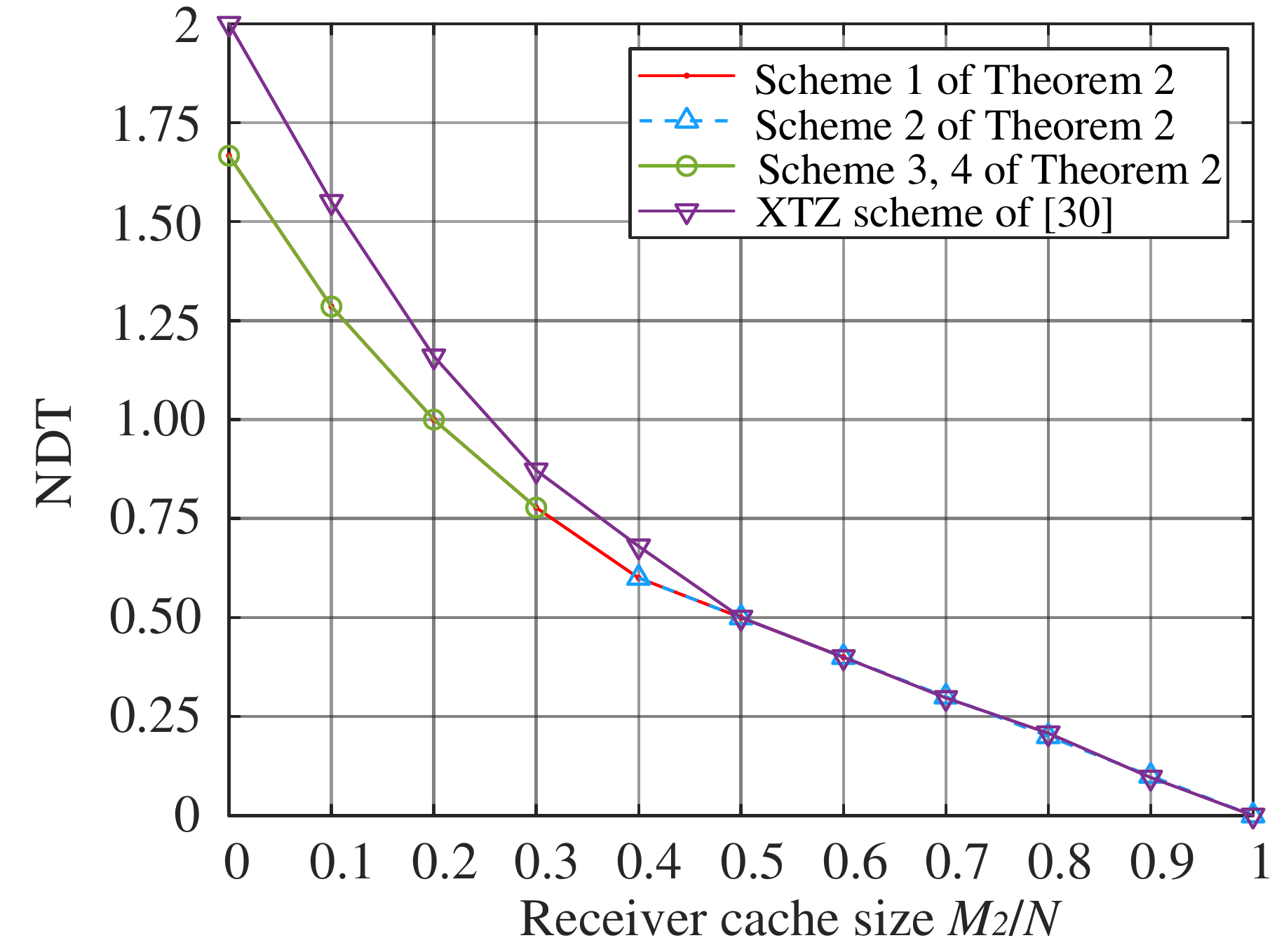}
	\caption{Achievable NDTs of  XTZ scheme in \cite{XTZ} and four schemes of Theorem \ref{PC-mapda} in the $15 \times 10$ linear network when $K=10$, $L=6$ and $\frac{M_{\text{T}}}{N}=\frac{1}{2}$ with $0\leq \frac{{M_{\text{U}}}}{N}\leq 1$ \label{fig-perfermance2}}
\end{figure}

By Table \ref{tab-relationship}, we can see that our schemes reduce the NDT compared to the XTZ scheme when $\frac{M_\text{U}}{N}+\frac{M_\text{T}}{N}
\frac{L}{K}\left\lceil \frac{K}{L} \right\rceil \geq 1$ or $\frac{M_\text{U}}{N}+\frac{M_\text{T}}{N}< 1$ and $L\nmid K$; when $\frac{M_\text{U}}{N}+\frac{M_\text{T}}{N}\geq 1$ our scheme has the same optimal NDT as the XTZ scheme. Let us also take a numerical comparison to verify our claim in Table  \ref{tab-relationship}.
When $\frac{M_{\text{T}}}{N}=\frac{1}{2}$, $K=10$, and $L=6$, we can obtain the NDTs of our schemes in Theorem \ref{PC-mapda} and the XTZ scheme respectively, as listed in Fig.\ref{fig-perfermance2}.
We can see that the NDTs of Scheme 1, Scheme 3, and Scheme 4 are smaller than the NDT of the XTZ scheme when $\frac{M_{\text{U}}}{N}\leq\frac{1}{2}$, which is corresponding to the conditions in the second and third row of Table \ref{tab-relationship}. Recall that the XTZ scheme already achieves the optimal NDT when
$\frac{1}{2}\leq\frac{M_{\text{U}}}{N}$ for the condition in first row of Table \ref{tab-relationship}. Clearly, Scheme 1 and Scheme 2 also have the same NDT which implies that they are also
optimal.

Finally, by Table \ref{tab-compare-2} we can see that there exist some schemes in Theorem \ref{PC-mapda} having both smaller NDT and subpacketizations than those of the XTZ schemes.
{
\begin{table*}[htbp]\small
	\caption{The subpacketizations of  XTZ scheme in \cite{XTZ} and our schemes in Theorem \ref{PC-mapda}.}
	\centering
	\renewcommand\arraystretch{1.2}{
		\setlength{\tabcolsep}{1.2mm}{
\begin{tabular}{|c|c|c|c|c|c|c|c|c|c|c|}
\hline
 \multicolumn{4}{|c|}{ } &\multicolumn{5}{|c|}{subpacketization}&\multicolumn{2}{|c|}{NDT}\\
 \cline{5-11}
\multicolumn{4}{|c|}{} & \multicolumn{1}{|c|}{XTZ scheme} &\multicolumn{4}{|c|}{Theorem $2$}& \multicolumn{1}{|c|}{XTZ scheme} &\multicolumn{1}{|c|}{Theorem $2$}  \\
\hline
$K$ & $M_{\text{T}}/N$ & $M_{\text{U}}/N$ & $L$ & \eqref{eq-subpacket-XTZ} & \text{Scheme 1}&  \text{Scheme 2}&\text{Scheme 3}&\text{Scheme 4}& \eqref{eq-Tao-ndt}& \\
\hline
 10 & 0.2 & 0.2 & 5 &25&50&$-$& 25 &100  &1.04 &2 \\
 \hline
 10 & 0.33 & 0.33 & 6 &180  &300 &$-$ &60  &180  &1 &1 \\
 \hline
 10 & 0.67 & 0.33 & 6 & 90 &420 & 60&$-$  &$-$  &0.6 &0.6 \\
 \hline
 10 & 0.33 & 0.4 & 6 & 180 &120 & $-$&60  &120  &0.84 &0.75 \\
 \hline
 50 & 0.1 & 0.1 & 10 &100  &200 & $-$&100  &900  &1.26 &4.5 \\
 \hline
 50 & 0.1 & 0.4 & 10 &2100  &200 & $-$&2100  &900  &0.66 &4.5 \\
 \hline
  100 & 0.2 & 0.2 & 50 & 8.7E$+$19 &500 & $-$& 2500 &1000  &2 &2 \\
 \hline
 100 & 0.4 & 0.2 & 50 & 2.8E$+$22 &750 &$-$ &750  &1500  &1.33 & 1.33\\
 \hline
 100 & 0.5 & 0.5 & 100 & 5.1E$+$30 &400 &5000 &$-$  &$-$  &0.5 & 0.5\\
 \hline
	\end{tabular}}}\label{tab-compare-2}
\end{table*}
}
\subsubsection{The placement strategy for transmitters of XTZ scheme in \cite{XTZ}}
\label{subsub-placement-transmitter}
For the placement strategy of the transmitters, when $\frac{L M_{\text{T}}}{N} = 1$ each of every $L$ consecutive transmitters caches a distinct part of each file. This is the same as our placement strategy for the transmitters. When $\frac{L M_{\text{T}}}{N}> 1$ the placement strategy for the transmitters relies on the placement strategy for the users. Clearly, our placement strategy for the transmitters is independent to the placement strategy for the users.

\subsubsection{The delivery strategy of XTZ scheme in \cite{XTZ}}
\label{subsub-delivery}	
In the delivery phase of the XTZ scheme, the computational complexity mainly comes from the design of the precoding matrices and decoding matrices to align or neutralize the interference in the communication\cite{XTZ}.  Specifically, it requires the users to first wait for multiple transmission slots and then compute the decoding matrices, which are the inverse of the multiplication of channel coefficients matrices and the precoding matrices. For example, for the case $L{M_{\text{T}}}/{N}=1$, the XTZ scheme applies the interference alignment in transmission that requires in total $L \binom{L-1}{\frac{M_{\text{U}}L}{N}} n^\rho+\binom{L-1}{\frac{M_{\text{U}}L}{N}+1})(n+1)^\rho$ transmission slots where $n\in\mathbb{Z}^+$ and $\rho=(K+L-1)(L-\frac{M_{\text{U}}L}{N}-1)$.  On the contrary, there exists no matrix computation in our schemes due to one-shot binary communication, i.e., the user can directly obtain the desired symbol after every transmission.

\subsection{Sketch of Construction in Theorem \ref{th-PCPDA-CCS}}
\label{subsect-sketch}
Let us consider a  $(K+L-1)\times K =7\times 5 $ partially connected linear network with  $(K,L,M_{\text{T}},M_{\text{U}},N)=(5,3,10,3,15)$. We will show how to generate a  coded caching scheme based on  $g$-$(r,K,F_1,Z_1,S_1)=5$-$(4,5,5,1,4)$ MAPDA $\mathbf{Q}$ in Example \ref{ex-1}. Clearly, $\frac{r}{\lceil K/L\rceil}=2$ is an integer, i.e., the condition in Theorem \ref{th-PCPDA-CCS} holds.

Our main construction idea is that we first generate a new $g$-$(r,K,LF_1,LZ_1,LS_1)=5$-$(4,5,15,3,12)$ MAPDA $\mathbf{P}$ based on $\mathbf{Q}$ and an $LF_1\times (K+L-1)=15\times 7$ array $\mathbf{T}$ called  transmitter caching array. Then, we propose the placement strategies for the transmitters and users according to the stars in $\mathbf{T}$ and $\mathbf{P}$, respectively. According to the integers in  $\mathbf{P}$, the delivery strategy is proposed. So, our method contains constructing two arrays $\mathbf{P}$ and $\mathbf{T}$, and realizing a scheme via $\mathbf{T}$ and $\mathbf{P}$.

\subsubsection{Constructing $\mathbf{P}$ and $\mathbf{T}$}
	By placing $\mathbf{Q}$ in Example \ref{ex-1} three times vertically and then increasing the integers in $\mathbf{Q}$ by the occurrence orders (from up to down) of $\mathbf{Q}$, we have a $15\times 5$ array $\mathbf{P}$  as follows.
	\begin{align}\label{ex-MAPDA-PC}\small
		\mathbf{P}=\left(\begin{array}{c}
			\mathbf{Q}  \\
			\mathbf{Q}+4 \\
			\mathbf{Q}+8
		\end{array}\right).
	\end{align} It is not difficult to check that the obtained array $\mathbf{P}$ is a $g$-$(r,K,LF_1,LZ_1,LS_1)=5$-$(4,5,15,3,12)$ MAPDA.
	
	Now let us construct the $15\times 7$ array $\mathbf{T}$. As illustrated in Fig.  \ref{fig-trans-cache},
{\begin{figure*}[http!]
\centering
		\includegraphics[scale=0.7]{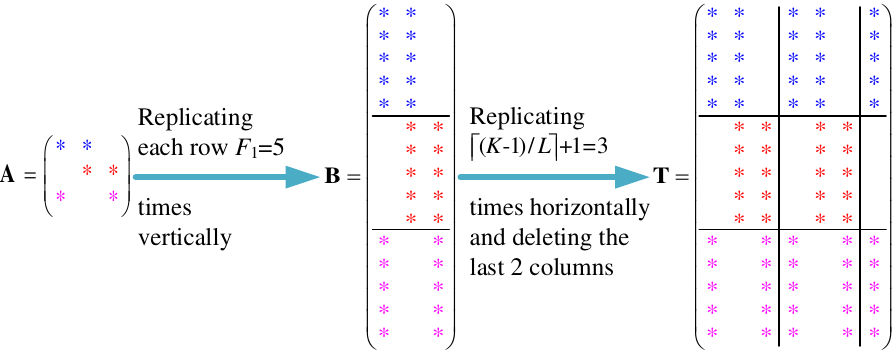}
		\caption{The flow diagram of constructing transmitter caching array $\mathbf{T}$\label{fig-trans-cache}}
\end{figure*}
}
we first construct an $L\times L=3\times 3$ square $\mathbf{A}$ in which each row has $t=LM_{\text{T}}/N=2$ cyclic stars, replicate each row of $\mathbf{A}$ vertically $F_1=5$ times to obtain $\mathbf{B}$, and further replicate $\mathbf{B}$ horizontally $\lceil (K-1)/L\rceil+1=3$ times, and finally delete the last  $L\cdot(\lceil (K-1)/L\rceil+1)-(K+L-1)=9-7$ columns to obtain our desired array $\mathbf{T}$ in Fig. \ref{fig-trans-cache}.

In order to simplify our introduction, we use the $(l, f)$ to represent the row label for each $l\in [L]$ and $ f\in [F_1]$. That is, the arrays $\mathbf{P}=\mathbf{P}((l, f),k)_{l\in [L], f\in [F_1],k\in [K]}$ and  $\mathbf{T}=\mathbf{T}((l, f),j)_{l\in [L],  f\in [F_1], j\in [K+L-1]}$.

\subsubsection{Generating a scheme via $\mathbf{T}$ and $\mathbf{P}$} Let the  $K+L-1$ columns and $LF_1$ rows of $\mathbf{T}$ in Fig. \ref{fig-trans-cache} represent the transmitters  and packets of each file, respectively; and let the columns $K$ and $LF_1$ rows of $\mathbf{P}$ in \eqref{ex-MAPDA-PC} represent the users and packets of each file, respectively. Then we can obtain a coded caching scheme for the $(K+L-1)\times K=7\times 5$ partially connected network as follows.

$\bullet$ {\textbf{Placement phase:}} We divide each file into $LF_1=15$ packets with equal size, i.e., for any $n\in[15]$, we have $W_n=\{W_{n,(l, f)} | l\in [L]=[3], f\in [F_1]=[5]\}$. Each transmitter and user caches the packets according to the stars in array $\mathbf{T}$ and $\mathbf{P}$, respectively. Specifically, each transmitter T$_{j}$ where $j\in [K+L-1]$ caches the packet $W_{n,(l, f)}$ if the entry $\mathbf{T}((l, f),j)=*$, i.e., all the transmitters cache the following packets.
\begin{align}
\mathcal{Z}_{\text{T}_1}=&\{W_{n,(1, f)}, W_{n,(3, f)} | f\in [5], n\in[15]\},\nonumber\\
\mathcal{Z}_{\text{T}_2}=&\{W_{n,(1, f)}, W_{n,(2, f)} |  f\in [5], n\in[15]\},\nonumber\\			
\mathcal{Z}_{\text{T}_3}=&\{W_{n,(2, f)}, W_{n,(3, f)} | f\in [5], n\in[15]\},\nonumber\\ 	
\mathcal{Z}_{\text{T}_4}=&\{W_{n,(1, f)}, W_{n,(3, f)} | f\in [5], n\in[15]\},\nonumber\\	
\mathcal{Z}_{\text{T}_5}=&\{W_{n,(1, f)}, W_{n,(2, f)} | f\in [5], n\in[15]\},\nonumber\\  	
\mathcal{Z}_{\text{T}_6}=&\{W_{n,(2, f)}, W_{n,(3, f)} | f\in [5], n\in[15]\},\nonumber\\			
\mathcal{Z}_{\text{T}_7}=&\{W_{n,(1, f)}, W_{n,(3, f)} | f\in [5], n\in[15]\}.\label{eq-Ex-cache-T}
\end{align} Each user U$_k$, for $k\in [K]$, caches the packet $W_{n,(l, f)}$ if the entry $\mathbf{P}((l, f),k)=*$, i.e., all the users cache the following packets.
\begin{align}
\mathcal{Z}_{\text{U}_1}=&\{W_{n,((l,1)} |  l\in [L]=[3], n\in[15]\},\nonumber\\
\mathcal{Z}_{\text{U}_2}=&\{W_{n,((l,2)} |  l\in [L]=[3], n\in[15]\},\nonumber\\
\mathcal{Z}_{\text{U}_3}=&\{W_{n,((l,3)} |  l\in [L]=[3], n\in[15]\}, \nonumber\\
\mathcal{Z}_{\text{U}_4}=&\{W_{n,((l,4)} |  l\in [L]=[3], n\in[15]\},\nonumber\\
\mathcal{Z}_{\text{U}_5}=&\{W_{n,((l,5)} |  l\in [L]=[3], n\in[15]\}.
\label{eq-Ex-cache-U}
		\end{align}

$\bullet$ {\textbf {Delivery phase:}} Assume that the request vector is $\textbf{d}=(1,2,3,4,5)$. For each block $s$, we send all the packets indexed by the integer $s$ in $\mathbf{P}$. For instance, when $s=1$ we have
\begin{align*}
\mathbf{P}((1,2),1)
&=\mathbf{P}((1,1),2)
=\mathbf{P}((1,1),3)\\
&=\mathbf{P}((1,1),4)
=\mathbf{P}((1,1),5)\\
&=1.
\end{align*}
Then, we will let the transmitters send the packets
\begin{align*}
\mathbf{W}^{(1)}
=(\tilde{W}_{1,(1,2)}, \tilde{W}_{2,(1,1)},\tilde{W}_{3,(1,1)},\tilde{W}_{4,(1,1)},\tilde{W}_{5,(1,1)})^{\top},
\end{align*}
where $\tilde{W}_{n,(l, f)}$ denotes the coded version of  any subfile ${W}_{n,(l, f)}$, for ${l\in[L], f\in [F_1]}$.
Clearly, these packets are required by all users. From \eqref{eq-Ex-cache-T}, transmitters T$_3$ and T$_6$ do not cache any packet of $W^{(1)}$. So the signals transmitted by all the transmitters can be represented as follows.
\begin{small}\begin{align*}\setlength{\arraycolsep}{3pt}
\mathbf{X}^{(1)}\!&=\!
		\left(
		\begin{array}{c}
			\mathbf{x}^{(1)}_{\text{T}_1}\\
			\mathbf{x}^{(1)}_{\text{T}_2}\\
			\vdots\\
			\mathbf{x}^{(1)}_{\text{T}_7}
		\end{array}
		\right)\nonumber\\
		&=\!\mathbf{V}^{(1)}\mathbf{W}^{(1)}\\
		&=
\left(
  \begin{array}{ccccc}
    v^{(1)}_{1,1} & v^{(1)}_{1,2} & v^{(1)}_{1,3} & v^{(1)}_{1,4} & v^{(1)}_{1,5} \\
    v^{(1)}_{2,1} & v^{(1)}_{2,2} & v^{(1)}_{2,3} & v^{(1)}_{2,4} & v^{(1)}_{2,5} \\
    0 & 0 & 0 & 0 & 0 \\
    v^{(1)}_{4,1} & v^{(1)}_{4,2} & v^{(1)}_{4,3} & v^{(1)}_{4,4} & v^{(1)}_{4,5} \\
    v^{(1)}_{5,1} & v^{(1)}_{5,2} & v^{(1)}_{5,3} & v^{(1)}_{5,4} & v^{(1)}_{5,5} \\
    0 & 0 & 0 & 0 & 0 \\
    v^{(1)}_{7,1} & v^{(1)}_{7,2} & v^{(1)}_{7,3} & v^{(1)}_{7,4} & v^{(1)}_{7,5}
  \end{array}
\right)
		\left(
		\begin{array}{c}
			\tilde{W}_{1,(1,2)}\\
           \tilde{W}_{2,(1,1)}\\
            \tilde{W}_{3,(1,1)}\\
            \tilde{W}_{4,(1,1)}\\
            \tilde{W}_{5,(1,1)}
		\end{array}
		\right).
		\label{eq:X in block-1}
\end{align*}
\end{small}Here each entry $v^{(1)}_{j,k}$, for $j\in \{1,2,4,5,7\}$, can be chosen from any value of $\mathbb{C}$. According to the  partially connected topology and from \eqref{eq-signal-partially}, the signals received by all the users are
\begin{small}
\begin{align*}
\setlength{\arraycolsep}{1pt}
\mathbf{Y}^{(1)}
=&
\setlength{\arraycolsep}{2pt}
	\left(
	\begin{array}{c}
		\mathbf {y}^{(1)}_{\text{U}_1}\\
		\mathbf {y}^{(1)}_{\text{U}_2}\\
		\mathbf {y}^{(1)}_{\text{U}_3}\\
		\mathbf {y}^{(1)}_{\text{U}_4}\\
		\mathbf {y}^{(1)}_{\text{U}_5}
	\end{array}\right)
=\left(
\begin{array}{ccccccc}
h^{(1)}_{1,1} & h^{(1)}_{1,2} & h^{(1)}_{1,3} &0 & 0 &0& 0\\
0 & h^{(1)}_{2,2} & h^{(1)}_{2,3} & h^{(1)}_{2,4} & 0&0&0 \\
0 & 0 & h^{(1)}_{3,3} & h^{(1)}_{3,4} & h^{(1)}_{3,5} & 0 & 0 \\
0 & 0 & 0& h^{(1)}_{4,4} & h^{(1)}_{4,5} & h^{(1)}_{4,6} & 0 \\
0& 0&0&0 &h^{(1)}_{5,5} & h^{(1)}_{5,6} & h^{(1)}_{5,7}
  \end{array}
\right)\\
&\left(
  \begin{array}{ccccc}
    v^{(1)}_{1,1} & v^{(1)}_{1,2} & v^{(1)}_{1,3} & v^{(1)}_{1,4} & v^{(1)}_{1,5} \\
    v^{(1)}_{2,1} & v^{(1)}_{2,2} & v^{(1)}_{2,3} & v^{(1)}_{2,4} & v^{(1)}_{2,5} \\
    0 & 0 & 0 & 0 & 0 \\
    v^{(1)}_{4,1} & v^{(1)}_{4,2} & v^{(1)}_{4,3} & v^{(1)}_{4,4} & v^{(1)}_{4,5} \\
    v^{(1)}_{5,1} & v^{(1)}_{5,2} & v^{(1)}_{5,3} & v^{(1)}_{5,4} & v^{(1)}_{5,5} \\
    0 & 0 & 0 & 0 & 0 \\
    v^{(1)}_{7,1} & v^{(1)}_{7,2} & v^{(1)}_{7,3} & v^{(1)}_{7,4} & v^{(1)}_{7,5}
  \end{array}
\right)		\left(
		\begin{array}{c}
			\tilde{W}_{1,(1,2)}\\
            \tilde{W}_{2,(1,1)}\\
            \tilde{W}_{3,(1,1)}\\
            \tilde{W}_{4,(1,1)}\\
            \tilde{W}_{5,(1,1)}
		\end{array}
		\right)\\
=&\setlength{\arraycolsep}{2pt}\left(
\begin{array}{ccccccc}
h^{(1)}_{1,1}& h^{(1)}_{1,2}& 0           & 0           &0\\
0            & h^{(1)}_{2,2}&h^{(1)}_{2,4}& 0           &0 \\
0            & 0            &h^{(1)}_{3,4}&h^{(1)}_{3,5}&0 \\
0            & 0            &h^{(1)}_{4,4}&h^{(1)}_{4,5}&0 \\
0            & 0            &0            &h^{(1)}_{5,5}&h^{(1)}_{5,7}
  \end{array}
\right)\left(
  \begin{array}{ccccc}
    v^{(1)}_{1,1} & v^{(1)}_{1,2} & v^{(1)}_{1,3} & v^{(1)}_{1,4} & v^{(1)}_{1,5} \\
    v^{(1)}_{2,1} & v^{(1)}_{2,2} & v^{(1)}_{2,3} & v^{(1)}_{2,4} & v^{(1)}_{2,5} \\
    v^{(1)}_{4,1} & v^{(1)}_{4,2} & v^{(1)}_{4,3} & v^{(1)}_{4,4} & v^{(1)}_{4,5} \\
    v^{(1)}_{5,1} & v^{(1)}_{5,2} & v^{(1)}_{5,3} & v^{(1)}_{5,4} & v^{(1)}_{5,5} \\
    v^{(1)}_{7,1} & v^{(1)}_{7,2} & v^{(1)}_{7,3} & v^{(1)}_{7,4} & v^{(1)}_{7,5}
  \end{array}
\right)\\
&\left(
		\begin{array}{c}
			\tilde{W}_{1,(1,2)}\\
            \tilde{W}_{2,(1,1)}\\
            \tilde{W}_{3,(1,1)}\\
            \tilde{W}_{4,(1,1)}\\
            \tilde{W}_{5,(1,1)}
		\end{array}
		\right)
=\mathbf{H}^{(1)}_1\mathbf{V}^{(1)}_1\mathbf{W}^{(1)}
=\mathbf{R}^{(1)}_1\mathbf{W}^{(1)}.
\end{align*}
\end{small}
It is not difficult to check that all the users can decode their requesting packets by their cached packets respectively if
\begin{eqnarray} 
\label{eq-resovle-1}
\mathbf{R}^{(1)}_1=
\left(
  \begin{array}{ccccc}
1 & a_1 & a_2 & a_3 & a_4\\
a_5 & 1 & 0 & 0 & 0\\
0 &  0  &  1  &  0  &  0\\
0 & 0 & 0 & 1 & 0\\
 0 & 0 & 0 & 0 & 1
  \end{array}
\right).
\end{eqnarray}
Here  $a_i$, for $i\in[5]$, can be any complex number in $\mathbb{C}$. For instance, given  $\mathbf{R}^{(1)}_1$ in \eqref{eq-resovle-1}, the user U$_1$ observes the coded signal
\begin{align*}
\mathbf {y}^{(1)}_{\text{U}_1}=&\tilde{W}_{1,(1,2)}+a_1 \tilde{W}_{2,(1,1)}+a_2\tilde{W}_{3,(1,1)}\\
&\ \ \ \ \ \ \ \ \ \ \ \ \ \ \ \ \  +a_3\tilde{W}_{4,(1,1)}+a_4\tilde{W}_{5,(1,1)}.
\end{align*}
From \eqref{eq-Ex-cache-U}, user U$_1$ has cached the packets
$W_{2,(1,1)}$, $W_{3,(1,1)}$, $W_{4,(1,1)}$ and $W_{5,(1,1)}$. So, it can obtain the  coded packets $\tilde{W}_{2,(1,1)}$, $\tilde{W}_{3,(1,1)}$, $\tilde{W}_{4,(1,1)}$ and $\tilde{W}_{5,(1,1)}$. Clearly it can decode the required   $\tilde{W}_{1,(1,2)}$ based on the received $\mathbf {y}^{(1)}_{\text{U}_1}$ and then recover the desired packet $W_{1,(1,2)}$.

Recall that the non-zero entry in  $\mathbf{H}^{(1)}$ is independent and identically distributed in $\mathbb{C}$. 
%From \eqref{eq-signal-partially}, $\mathbf{H}^{(1)}$ is full rank.
For instance, let
\begin{align*}
\mathbf{H}^{(1)}_1\!=\!\left(
  \begin{array}{ccccc}
     1  &   2  &   0   &  0 &    0\\[-0.15cm]
     0  &   1 &    2 &    0  &   0\\[-0.15cm]
     0  &   0  &   1  &   2  &   0\\[-0.15cm]
     0  &   0  &   1 &    3  &   0\\[-0.15cm]
     0  &   0   &  0  &   1  &   2
  \end{array}
\right),
\mathbf{R}^{(1)}_1\!=\!\left(
  \begin{array}{ccccc}
1 & 1 & 1 & 1 & 1\\[-0.15cm]
1 & 1 & 0 & 0 & 0\\[-0.15cm]
0 &  0  &  1  &  0  &  0\\[-0.15cm]
0 & 0 & 0 & 1 & 0\\[-0.15cm]
 0 & 0 & 0 & 0 & 1
  \end{array}
\right).
\end{align*} We have $\mathbf{H}^{(1)}$ is full rank and
\begin{align*}
\left(\mathbf{H}^{(1)}_1\right)^{-1}=\left(
  \begin{array}{ccccc}
         1  & -2       &  12     &  -8       &       0\\[-0.15cm]
         0  &  1       & -6      &   4       &       0\\[-0.15cm]
         0  &       0  &  3      &  -2       &       0\\[-0.15cm]
         0  &       0  & -1      &   1       &       0\\[-0.15cm]
         0  &       0  &  0.5    &  -0.5     &  0.5
  \end{array}
\right).
\end{align*} From \eqref{eq-resovle-1} we have
\begin{eqnarray*}
\mathbf{V}^{(1)}_1&=\left(\mathbf{H}^{(1)}_1\right)^{-1} \left(
  \begin{array}{ccccc}
1 & 1 & 1 & 1 & 1\\[-0.15cm]
1 & 1 & 0 & 0 & 0\\[-0.15cm]
0 &  0  &  1  &  0  &  0\\[-0.15cm]
0 & 0 & 0 & 1 & 0\\[-0.15cm]
 0 & 0 & 0 & 0 & 1
  \end{array}
\right)\\
&=
\left(
  \begin{array}{ccccc}
   -1       & -1      &  13     &  -7     &   1 \\[-0.15cm]
    1       &  1      &  -6     &  4      &        0\\[-0.15cm]
         0  &       0 &   3     &  -2     &        0\\[-0.15cm]
         0  &       0 & -1      &  1      &       0\\[-0.15cm]
         0  &       0 &  0.5    &  -0.5   &   0.5
  \end{array}
\right).
\end{eqnarray*}Then, we have the precoding matrix
\begin{eqnarray*}
\mathbf{V}^{(1)}=\left(
  \begin{array}{ccccc}
   -1       & -1      &  13     &  -7     &   1 \\[-0.15cm]
    1       &  1      &  -6     &  4      &        0\\[-0.15cm]
    0 & 0 & 0 & 0 & 0 \\[-0.15cm]
         0  &       0 &   3     &  -2     &        0\\[-0.15cm]
         0  &       0 & -1      &  1      &       0\\[-0.15cm]
    0 & 0 & 0 & 0 & 0 \\[-0.15cm]
         0  &       0 &  0.5    &  -0.5   &   0.5
  \end{array}
\right).
\end{eqnarray*}
So, for any give channel matrix $\mathbf{H}^{(1)}$, we can always obtain a precoding matrix $\mathbf{V}^{(1)}$ such that \eqref{eq-resovle-1} holds. Similarly, we can check the other $s=2$, $\ldots$, $12$.

From \eqref{eq-comput-NDT} and \eqref{eq-def-DoF}, we can obtain the NDT  $\tau_{\text{new}}(M_{\text{T}}=10,M_{\text{U}}=3)$ and sum-DoF$_{new}$ as follows.
\begin{align*}
\tau_{\text{new}}(M_{\text{T}}&=10,M_{\text{U}}=3)
=\frac{LS_1}{LF_1}
 =\frac{12}{15}=\frac{4}{5},\\
\text{Sum-DoF}_{new} &= \frac{K(1-M_\text{U}/N)}{\tau_{\text{new}}(M_{\text{T}}=10,M_{\text{U}}=3)} =5.
\end{align*}

Now let us see the performance of the XTZ scheme. By Lemma \ref{le-Tao} we have to use the memory sharing to obtain the $(K,L,M_{\text{T}},M_{\text{U}},N)=(5,3,10,3,15)$
 coded caching scheme for the partially connected linear network generated by the $(K=5,L=3,M_\text{T}^{'}=10,M_\text{U}^{'}=0,N=15)$ scheme, say Scheme A, where each file has $\frac{2}{5}V$ bits, and the $(K=5,L=3,M_\text{T}^{''}=10,M_\text{U}^{''}=5,N=15)$ scheme, say Scheme B, where each file has $\frac{3}{5}V$ bits, in \cite{XTZ} respectively, since
\begin{align*}
&M_\text{T}^{'}\cdot\frac{2V}{5}+M_\text{T}^{''}\cdot\frac{3V}{5}
=10\cdot V=M_\text{T},\\
&M_\text{U}^{'}\cdot\frac{2V}{5}+M_\text{U}^{''}\cdot\frac{3V}{5}
=5\cdot\frac{3V}{5}=3V=M_\text{U}.
\end{align*}
By Lemma \ref{le-Tao}, we have $p'=\frac{M_{\text{T}}^{'} L}{N}=2$ and $q'=\frac{M_{\text{U}}^{'} L}{N}=0$. Then, by third statement of Lemma \ref{le-Tao}, the NDT of Scheme A is
\begin{align*}
	\tau_{\text{A}}(M_{\text{T}}^{'} = 10, M_{\text{U}}^{'}=0)=\frac{L-\frac{M_{\text{U}}^{'} L}{N}}{\frac{M_{\text{T}}^{'} L}{N}+\frac{M_{\text{U}}^{'} L}{N}}= \frac{3}{2}.
\end{align*}
Similarly, we have $p^{''}=\frac{M_{\text{T}}^{''} L}{N}=2$ and $q^{''}=\frac{M_{\text{U}}^{''} L}{N}=1$. Then, by the second statement of Lemma \ref{le-Tao}, the NDT of Scheme B is
\begin{align*}
	\tau_{\text{A}}(M_{\text{T}}^{''}=10,M_{\text{U}}^{''}=5)
	=\frac{L-\frac{M_{\text{U}}^{''} L}{N}}{L}
	=\frac{2}{3}.
\end{align*}Thus, the obtained scheme with $M_\text{U}/N=1/5$ has the following NDT
\begin{align*}
&\tau_{\text{XT}}(M_{\text{T}}=10,M_{\text{U}}=3)\\
&=
\frac{2}{5}\cdot\tau_{\text{A}}(M_{\text{T}}=10,M_{\text{U}}^{'}=0)
	+\frac{3}{5}\cdot \tau_{\text{B}}(M_{\text{T}}=10,
	M_{\text{U}}^{''}=5)\\
	&=\frac{2}{5}\cdot\frac{3}{2}
	+\frac{3}{5}\cdot\frac{2}{3}= 1 > \frac{4}{5}=\tau_{\text{new}}(M_{\text{T}}=10,M_{\text{U}}=3),
\end{align*}
and the following Sum-DoF
\begin{align*}
	\text{Sum-DoF}_{X}&=
	\frac{K(1-M_\text{U}/N)}{\tau_{\text{XT}}
		(M_{\text{XT}}=10,M_{\text{U}}=3)} \\ &=4<5=\text{Sum-DoF}_{new}.
\end{align*}
So, our scheme achieves a smaller NDT and a larger Sum-DoF  than that of the XTZ scheme.
\section{The proof of Theorem \ref{th-PCPDA-CCS}}
\label{sec-proof-main-result}
Let us consider the $(K,r,M_{\text{T}},M_{\text{U}},N)$ partially connected coded caching problem, where $t=\frac{LM_{\text{T}}}{N}\in[0:L]$ and $z=\frac{KM_{\text{U}}}{N}\in [0:K]$. Given a $(r,K,F_1,Z_1,S_1)$ MAPDA $\mathbf{Q}$ satisfying $\frac{r}{\lceil K/L\rceil}=t$, as introduced in Subsection \ref{subsect-sketch}, we first construct a $(r, K,L F_1,L Z_1,L S_1)$ MAPDA $\mathbf{P}$ and an $LF_1\times (K+L-1)$  transmitter caching array $\mathbf{T}$, and then generate our desired partially connected coded caching scheme by using these two arrays.
\subsection{Constructing MAPDA $\mathbf{P}$ and Transmitter Cache Array $\mathbf{T}$}
We can obtain a new array $\mathbf{P}$ by replicating $\mathbf{Q}$ vertically $L$ times and then increasing the integers in $\mathbf{Q}$ by the occurrence orders (from up to down) of $\mathbf{Q}$, i.e., the $LF_1\times K$ array $\mathbf{P}$ is constructed as
\begin{eqnarray*}
\label{eq-cons-P}
\mathbf{P}=
\!\left(\mathbf{P}((l, f),k)\!\right)_{l\in[L],\! f\in [F],\!k\in [K]}
=\!\left(\begin{array}{c}
				\mathbf{Q}  \\
				\mathbf{Q}+S \\
                  \vdots    \\
				\mathbf{Q}+(L-1)S
			\end{array}\!\right).
\end{eqnarray*}
It is easy to check that $\mathbf{P}$ is a $(r, K,L F_1,L Z_1,L S_1)$ MAPDA. From \eqref{eq-cons-P} each entry of $\mathbf{P}$ can be defined in the following way.
\begin{IEEEeqnarray}{rCl}\label{eq-define-P}
&&\mathbf{P}((l, f),k)\nonumber\\
&&\quad\quad=\mathbf{Q}( f,k)+(l-1)S_1, l\in[L], f\in [F_1], k\in [K].\quad\quad
\end{IEEEeqnarray}
 Here the sum $a+*=*$ for any integer $a\in \mathbb{Z}$.

Let us introduce the construction of $\mathbf{T}$. We first construct a cyclic star placement array that is defined in \cite{WCC} as follows.
\begin{definition}(Cyclic star placement)
\label{def-cyclic}
An $(L,t)$ star placement array $\mathbf{A}=\mathbf{A}(l,l')_{l,l'\in [L]}$ including stars and null entries, is referred to as a {\it cyclic star placement} array, if the stars in each row are placed in a cyclic wrap-around topology, i.e., each entry
\begin{equation*}
%\label{eq-cyclic}
\mathbf{A}(l,l')=*\ \text{only if}\ \ l'\in\{<l+\mu>_L| \mu\in[0:t-1]\}.
\end{equation*} 
%\begin{align}\label{eq-cyclic}
%\mathbf{A}(l,l')=*,\ \text{only if}\ \ l'\in\{<l+\mu>_L \ |\ \mu\in[0:t-1] \}.
%\end{align}
\end{definition}For instance, we can check that the following array
\begin{eqnarray*}
\mathbf{A}=\left(
  \begin{array}{ccc}
    * & * &  \\
      & * & * \\
       * &    & *
  \end{array}
\right)
\end{eqnarray*}is a
 $(3,2)$ star placement array which is listed in Fig. \ref{fig-trans-cache}. As illustrated in Fig. \ref{fig-trans-cache}, we replicate each row of $\mathbf{A}$ vertically $F_1$ times to obtain $\mathbf{B}$, and further replicate $\mathbf{B}$ horizontally $\lceil (K-1)/L\rceil+1=3$ times and delete  the last  $L\cdot(\lceil (K-1)/L\rceil+1)-(K+L-1)$ columns to obtain our desired array $\mathbf{T}$. In fact, each entry of the obtained array $\mathbf{T}=\mathbf{T}((l, f),j)_{l\in [L],f\in[F],j\in[K+L-1]}$ can be defined as follows.
\begin{eqnarray}\label{eq-placement-trans-array}
\mathbf{T}((l, f),j)=\left\{
\begin{array}{cc}
  * & \text{if}\  <j>_L \in\ \ \ \ \   \\
  &\{<l+\mu>_L | \mu\in[0:t-1]\}, \\
  \text{Null} & \text{otherwise}.
\end{array}
\right.
\end{eqnarray} For instance, when $t=2$, $K=5$, $F_1=5$,  $L=3$ and from \eqref{eq-placement-trans-array}, we have the transmitter caching array $\mathbf{T}$ listed in Fig. \ref{fig-trans-cache}.

\subsection{The Scheme Realized by $\mathbf{P}$ and $\mathbf{T}$}
Given a MAPDA $(r,K,LF_1,LZ_1,LS_1)$ $\mathbf{P}$ and a transmitter cache array $\mathbf{T}$ constructed in the above subsection, we can obtain a $LF$-division coded caching scheme for the $(K,L,M_{\text{T}},M_{\text{U}},N)$ for the $(K+L-1)\times K$ partially connected network in the following way.

$\bullet$  {\bf Placement phase:} Each file $W_n$ where $n\in [N]$ is divided into $LF_1$ packets with equal size, i.e., $W_{n}=(W_{n,(l, f)})_{l\in[L], f\in [F_1]}$. From \eqref{eq-placement-trans-array}, each transmitter T$_{j}$ where $j\in[K+L-1]$ caches the following packets.
\begin{eqnarray}
\label{eq-transmitter-caching-content}
\mathcal{Z}_{\text{T}_{j}}=\{W_{n,(l, f)} |\!\!\!\!\!\!&\mathbf{T}((l, f),j)=*,\ \ \ \ \ \ \ \ \ \ \ \nonumber\\
&l\in [L],\  f\in [F_1],\ n\in [N]\}.
\end{eqnarray} We can check that transmitter T$_{j}$ caches exactly $tF_1N$ packets. Recall that $t=\frac{LM_{\text{T}}}{N}$. So we have $M_{\text{T}}=\frac{tF_1N}{LF_1}$. From \eqref{eq-cons-P},  each user U$_k$ caches the following packets.
\begin{eqnarray}\label{eq-receier-caching-content}
\mathcal{Z}_{\text{U}_k}=\{W_{n,(l, f)} |\!\!\!\!\!\!&\mathbf{P}((l, f),k)=*,\ \ \ \ \ \ \ \ \ \ \ \nonumber\\
&l\in [L],\  f\in [F_1],\ n\in [N]\}.
\end{eqnarray}
We can check that user U$_k$ caches exactly $LZ_1N$ packets and $M_{\text{U}}=\frac{LZ_1N}{LF_1}=\frac{Z_1N}{F_1}$.

$\bullet$  {\bf Delivery phase:} For any request vector ${\bf d}$, the delivery strategy consists of $LS_1$ blocks. For each block $s\in [LS_1]$, we assume that there are $r_s$  entries $\mathbf{P}((l_{1}, f_{1}),k_1)$, $\mathbf{P}((l_{2}, f_{2}),k_2)$, $\ldots$ , $\mathbf{P}((l_{r_s}, f_{r_s}),k_{r_s})$ equal to $s$, where $l_{i}\in[L]$, $ f_{i}\in[F_1]$ and ${k}_{i}\in[K]$  for each $i\in [r_s]$. The vector of  packets to be transmitted in block $s$ and the user index set to recover these packets are denoted by
\begin{equation*}
			\label{eq-packet-user-time-s}
			\mathbf{W}^{(s)}=\left(
			\begin{array}{c}
				\tilde{W}_{d_{{k}_{1}},(l_{1}, f_{1})}\\
				\tilde{W}_{d_{{k}_{2}},(l_{2}, f_{2})}\\
				\vdots\\
				\tilde{W}_{d_{{k}_{r_s}},(l_{r_s}, f_{r_s})}
			\end{array}
			\right)\ \text{and} \ \mathcal{K}_s=\{{k}_{1},\!{k}_{2},\!\ldots,\!{k}_{r_s}\},
\end{equation*}
		respectively, where $\tilde{W}_{n,(l, f)}$ denotes the coded version of    ${W}_{n,(l, f)}$, for ${l\in[L], f\in [F_1]}$. By the third property C3 of the MAPDA, we have that $|\mathcal{K}_s|=r_s$. Without loss of generality, we assume that $k_1<k_2<\cdots<k_{r_s}$ and each user U$_{k_i}$ requires the packet $W_{d_{{k}_{i}},(l_{i}, f_{i})}$ where $i\in [r_s]$. Then, each transmitter T$_{j}$, $j \in [K+L-1]$ will transmit
\begin{eqnarray}\label{eq-delivery-transmiter-k-subset}
\mathbf{x}^{(s)}_{\text{T}_{j}}=\sum_{i=1}^{r_s} v^{(s)}_{j,i}\tilde{W}_{d_{{k}_{i}},(l_{i}, f_{i})},
\end{eqnarray}
where $v_{j,i}=0$ if $l_i\not\in \{<j>_L, <j-1>_L, \ldots, <j-t+1>_L\}$ by \eqref{eq-transmitter-caching-content} and \eqref{eq-placement-trans-array}, otherwise $v_{j,i}$ can be chosen any complex number.
For the users in $\mathcal{K}_s$, all the signals transmitted by transmitters at block $s$ are
\begin{eqnarray*}
%\label{eq-all-transmitted-signals}
\begin{split}
\mathbf{Y}(s)&=\mathbf{H}^{(s)}\mathbf{X}(s)
=\mathbf{H}^{(s)}\mathbf{V}^{(s)}\mathbf{W}^{(s)} \\
			&=\mathbf{H}^{(s)}\left({\bf v}^{(s)}_{1},{\bf v}^{(s)}_{2},\ldots,{\bf v}^{(s)}_{r_s}\right)
			\mathbf{W}^{(s)} \\
			&=\left(\mathbf{H}^{(s)}{\bf v}^{(s)}_{1},\mathbf{H}^{(s)}{\bf v}^{(s)}_{2},\ldots,\mathbf{H}^{(s)}{\bf v}^{(s)}_{r_s} \right)
			\mathbf{W}^{(s)} \\
			&=\mathbf{R}^{(s)}\mathbf{W}^{(s)}.
\end{split}
\end{eqnarray*}
Recall that each user  U$_{k_i}$ can receive the signal consisting of $\mathbf{x}^{(s)}_{\text{T}_{k_i}}$, $\mathbf{x}^{(s)}_{\text{T}_{k_i+1}}$, $\ldots$, $\mathbf{x}^{(s)}_{\text{T}_{k_i+L-1}}$ sent from the transmitters T$_{k_i}$, T$_{k_i+1}$, $\ldots$, T$_{k_i+L-1}$, respectively, i.e., the following signal from \eqref{eq-delivery-transmiter-k-subset}
\begin{eqnarray}
\begin{split}
{\bf y}_{k_i}^{(s)}
&=\sum\limits_{j=k_i}^{k_i+L-1}h^{(s)}_{k_i,j}
\mathbf{x}^{(s)}_{\text{T}_{j}}\\
&=\sum\limits_{j=k_i}^{k_i+L-1}h^{(s)}_{k_i,j}
\left(\sum_{i'=1}^{r_s} v^{(s)}_{j,i'}\tilde{W}_{d_{{k}_{i'}},(l_{i'}, f_{i'})}\right)\\
&=\sum\limits_{j=k_i}^{k_i+L-1}
\sum_{i'=1}^{r_s}h^{(s)}_{k_i,j} v^{(s)}_{j,i'}\tilde{W}_{d_{{k}_{i'}},(l_{i'}, f_{i'})}\\
&=\sum_{i'=1}^{r_s} \left(\sum\limits_{j=k_i}^{k_i+L-1} h^{(s)}_{k_i,j}v^{(s)}_{j,i'}\right)\tilde{W}_{d_{{k}_{i'}},(l_{i'}, f_{i'})}.
\end{split}
\label{eq-delivery-users-k-all-subset-1}
\end{eqnarray} 

In our scheme, we design the beamforming vector   $\{v_{j,i}\}$ to ensure one-shot delivery, i.e., each   $\mathbf{x}^{(s)}_{\text{T}_{j}}$ can be directly decoded by the desired users. We will explain the design of $\{v_{j,i}\}$ in the following subsection. 
\subsection{Decodability for Each User}

Now let us consider the subarray $\mathbf{P}^{(s)}$ generated by the rows $(l_{1}, f_{1})$, $(l_{2}, f_{2})$, $\ldots$, $(l_{r_s}, f_{r_s})$ and columns in $\mathcal{K}_s$. In the following, we will take the column indices in $\mathcal{K}_s$ and row indices $(l_{1}, f_{1})$, $(l_{2}, f_{2})$, $\ldots$, $(l_{r_s}, f_{r_s})$ as the columns indices and row indices of the subarray $\mathbf{P}^{(s)}$, respectively. For each $i\in [r_s]$, assume that there are $\lambda^{(s)}_{i}$ columns with indices in  $\mathcal{K}_s$ containing integers at the row $(l_{i}, f_{i})$ of $\mathbf{P}$. The set of these column indices can be written as
		\begin{eqnarray}
			\label{eq-not-cached}
			\mathcal{P}^{(s)}_{i}=\left\{k_{i'}\in \mathcal{K}_s\ |\ \mathbf{P}((l_{i}, f_{i}),{k}_{i'})\in [LS_1],\ i'\in [r_s]\right\}.
		\end{eqnarray}
Clearly, $|\mathcal{P}^{(s)}_{i}|=\lambda^{(s)}_{i}$. By \eqref{eq-receier-caching-content} and \eqref{eq-not-cached}, the demanded packet $W_{d_{{k}_{i}},(l_{i}, f_{i})}$ required by user U$_{{k}_{i}}$ is not cached by user U$_{{k}_{i'}}$ if  ${k}_{i'}\in\mathcal{P}^{(s)}_{i}$. So, \eqref{eq-delivery-users-k-all-subset-1} can be written as
\begin{align}\label{eq-delivery-users-k-subset}
{\bf y}_{k_i}^{(s)}&=\sum_{i'=1}^{r_s} \left(\sum\limits_{j=k_i}^{k_i+L-1} h^{(s)}_{k_i,j}v^{(s)}_{j,i'}\right)\tilde{W}_{d_{{k}_{i'}},(l_{i'}, f_{i'})}\nonumber\\
&=\underbrace{\left(\sum_{j=k_i}^{k_i+L-1} h^{(s)}_{k_i,j}v^{(s)}_{j,i}\right)\tilde{W}_{d_{{k}_{i}},(l_{i}, f_{i})}}_{\text{Required} \  \& \ \text{Uncaching packet}}
\\
&+\underbrace{\sum\limits_{i'\in[r_s]:k_{i}\in \mathcal{P}^{(s)}_{i'}\setminus\{k_{i'}\} }
\left(\sum_{j=k_i}^{k_i+L-1} h^{(s)}_{k_i,j}v^{(s)}_{j,{i'}}\right)\tilde{W}_{d_{{k}_{i'}},(l_{i'}, f_{i'}) }}_{\text{Unrequired}\  \& \ {\text{Uncaching packets}}}\nonumber\\
&+\underbrace{\sum\limits_{i'\in[r_s]:  k_{i}\in \mathcal{K}_s\setminus\mathcal{P}^{(s)}_{i'}} \left(\sum_{j=k_i}^{k_i+L-1} h^{(s)}_{k_i,j}v^{(s)}_{j,{i'}}\right)\tilde{W}_{d_{{k}_{i'}},(l_{i'}, f_{i'})}}
_{\text{Caching packets}},\nonumber
\end{align}
where the packet in the first term of the right side is required by user U$_{k_i}$; the packets in the second term of the right side are neither required nor cached by user U$_{k_i}$; the packets in the third term of the right side are not required but cached by user U$_{k_i}$. Clearly, we only need to consider the packets in the first two terms in  \eqref{eq-delivery-users-k-subset} since the user U$_{k_i}$ can cancel all the packets in the third term by its cached contents. In order to decode the desired packet $\tilde{W}_{d_{{k}_{i}},(l_{i}, f_{i})}$, we have to cancel the interfering packets in the second term of  \eqref{eq-delivery-users-k-subset}. Clearly, each user U$_{k_i}$ where $i\in [r_s]$ can decode its required packet $\tilde{W}_{d_{{k}_{i}},(l_{i}, f_{i})}$ by its received signal ${\bf y}_{k_i}^{(s)}$ and its cached packets $\mathcal{Z}_{\text{U}_{k_i}}$ if the following condition holds for any two different integers $i,i'\in [r_s]$.
\begin{align}
\label{eq-coding-vector2}
\left\{\begin{array}{c}
1=\sum_{j=k_i}^{k_i+L-1} h^{(s)}_{k_i,j}v^{(s)}_{j,i}=
\mathbf{H}^{(s)}(k_{i},\cdot)\mathbf{v}^{(s)}_{i}
=\mathbf{R}^{(s)}(i,i) ,\ \ \ \ \ \ \ \ \ \ \ \ \ \ \ \ \ \ \ \ \ \ \ \ \ \
\\[0.2cm]
\text{\tabincell{c}{$0=\sum_{j=k_i}^{k_i+L-1} h^{(s)}_{k_i,j}v^{(s)}_{j,{i'}}=
\mathbf{H}^{(s)}(k_{i},\cdot)\mathbf{v}^{(s)}_{i'}
=\mathbf{R}^{(s)}(i,i')$,\ \ \ \ \ \ \ \ \ \ \ \ \ \ \ \ \ \ \ \ \ \ \ \ \ \  \\
\ \ \ \ \ \ \ \ \ \ \ \ \ \ \ \ \ \ \ \ \ \ \ \ \ \ $k_{i}\in \mathcal{P}^{(s)}_{i'}\setminus\{k_{i'}\}$.}}
\end{array}\right.
\end{align} It is worth noting that the first equality in \eqref{eq-coding-vector2} means that the user U$_{k_i}$ can decode its desired  packet $\tilde{W}_{d_{{k}_{i}},(l_{i}, f_{i})}$, and the second equality in \eqref{eq-coding-vector2} means that the user U$_{k_i}$ can cancel its un-required and un-cached coded packet $\tilde{W}_{d_{{k}_{i'}},(l_{i'}, f_{i'})}$.
 Recall that for any $i\in [r_s]$ and $j\in [K+L-1]$, each coefficient $h^{(s)}_{k_i,j}=0$ if $j\not\in \{k_i,k_i+1,\ldots,k_i+L-1\}$, otherwise $h^{(s)}_{k_i,j}$ can be chosen any complex number in i.i.d distribution. Since all the required packets are sent by the transmitters in the whole communication process, each user can decode its required file if \eqref{eq-coding-vector2} holds for all $s\in [LF_1]$. So, it is sufficient to show that there exists a precoding matrix $\mathbf{V}^{(s)}$ satisfying \eqref{eq-coding-vector2}, for each $s\in [LS_1]$.
\subsection{The Existence of Precoding Matrix $\mathbf{V}^{(s)}$ Satisfying \eqref{eq-coding-vector2}}
Now we will show that we can choose appropriate coefficients $v_{j,i}$ for all  $j\in[K+L-1]$ and $i\in[r_s]$ such that \eqref{eq-coding-vector2} always holds. Recall that the transmitters cache all the packets in a successive placement from \eqref{eq-placement-trans-array}. So, there are exactly $t$ transmitters in $\{\text{T}_{k_i}, \text{T}_{k_i+1},\ldots, \text{T}_{k_i+L-1}\}$ caching  each packet $W_{d_{{k}_{i}},(l_{i}, f_{i})}$ where $i\in [r_s]$. Without loss of generality, we assume that the index set of the transmitters, each of which caches the $W_{d_{{k}_{i}},(l_{i}, f_{i})}$, is
\begin{equation}
\label{eq-T-cached-packet}
\mathcal{T}=\left\{i+mL\ \Big|\  i\in [t], m\in \left[\left\lfloor\frac{K+L-1}{L}\right\rfloor\right]\right\}
\bigcap[K+L-1].
\end{equation}
By the assumption that each coefficient $v^{(s)}_{j,i}=0$ in \eqref{eq-delivery-transmiter-k-subset} for each $j\not\in \mathcal{T}$, there are exactly $|\mathcal{T}|$ coefficients that can get any compplex number. So, \eqref{eq-coding-vector2} can be written as follows.
\begin{eqnarray}
\label{eq-vector-existence}
\mathbf{H}^{(s)}(\mathcal{P}^{(s)}_{i}, \mathcal{T})\mathbf{v}^{(s)}_{i}(\mathcal{T})={\bf b}=(b_1,b_2,\ldots, b_{\lambda^{(s)}_{i}})^{\top},
\end{eqnarray}where $b_{i'}=1$ if $k_{i'}=k_i$ and $b_{i'}=0$ if $k_{i'}\in \mathcal{P}^{(s)}_{i}\setminus\{k_i\}$.

Recall that $\frac{r}{\lceil K/L\rceil}=t$, i.e., $r=t\lceil K/L\rceil$. By the fourth condition in \eqref{C4} of Definition \ref{def-MAPDA}, we have $|\mathcal{T}|\geq t\lfloor\frac{K+L-1}{L}\rfloor\geq t\lceil\frac{K}{L}\rceil \geq \lambda^{(s)}_i=|\mathcal{P}^{(s)}_{i}|$. This implies that the number of rows of  $\mathbf{H}^{(s)}(\mathcal{P}^{(s)}_{i}, \mathcal{T})$ is less than or equal to the number of columns of $\mathbf{H}^{(s)}(\mathcal{P}^{(s)}_{i}, \mathcal{T})$. Furthermore, by Schwartz-Zippel Lemma \cite{DL}, we obtain the following result whose proof is included in Appendix \ref{sec:le-full-rank}.
\begin{lemma}
\label{le-subarray-full-rank}
$\mathbf{H}^{(s)}(\mathcal{P}^{(s)}_{i}, \mathcal{T})$ is a full row rank matrix.
\end{lemma}

By the linear algebra, there must exist a $\mathbf{v}^{(s)}_{i}(\mathcal{T})\in \mathbb{C}^{|\mathcal{T}|}$ satisfying \eqref{eq-vector-existence}. Then, by adding $K+L-1-|\mathcal{T}|$ zero entries in $\mathbf{v}^{(s)}_{i}(\mathcal{T})$, we can obtain a column vector $\mathbf{v}^{(s)}_{i}$ that satisfies \eqref{eq-coding-vector2}.

Finally, from \eqref{eq-comput-NDT} and \eqref{eq-def-DoF}, we can obtain the NDT  $\tau_{\text{new}}(M_{\text{T}},M_{\text{U}})=S/F$ and\begin{align*}
\text{sum-DoF}_{new}&=
\frac{K(1-M_{\text{U}}/N)}{\tau_{\text{new}}(M_{\text{T}},M_{\text{U}})}\\
&=K(1-\frac{Z_1}{F_1})\frac{F}{S}=g.
\end{align*}

\section{Conclusion}\label{sec-conclu}
  In this paper, we studied the coded caching problem  for the $(K,L,M_{\text{T}},M_{\text{U}},N)$ partially connected linear network. Firstly,  we showed that MAPDA can be also used to design the scheme for the partially connected linear network with a delicate design on the data placement on the transmitters and users. Consequently, by the existing MAPDAs and a delicate construction method, we obtain some new schemes for the partially connected linear network which have smaller NDT than that of the XTZ scheme for many cases. Furthermore,  our schemes operate in one-shot linear delivery and can significantly reduce the subpacketizations compared to the XTZ scheme. This implies that our schemes are communication-efficient  and have a wider range of applications and lower complexity of implementation.

\begin{appendices}
\section{Proof of Lemma~\ref{le-subarray-full-rank}}
\label{sec:le-full-rank}
First, the following notation and assumption can be used to simplify our introduction. Let $\mathbf{H}=\mathbf{H}^{(s)}(\mathcal{P}^{(s)}_{i}, \mathcal{T})$. We also use the row indices and column indices of $\mathbf{H}^{(s)}$ as the row indices and column indices of $\mathbf{H}$. That is,  $\mathbf{H}=(\mathbf{H}(k,j))_{k\in \mathcal{P}^{(s)}_{i},j\in \mathcal{T}}$. Without loss of generality, we assume that $\mathcal{P}^{(s)}_{i}=\{k_1,k_2,\ldots,k_{\lambda^{(s)}_{i}}\}$. The main idea is that we first find a non-zero path
$$(\mathbf{H}(k_1,j_1), \mathbf{H}(k_2,j_2),\ldots,\mathbf{H}(k_{\lambda^{(s)}_{i}},j_{\lambda^{(s)}_{i}})),$$ where $\mathbf{H}(k_{\lambda},j_{\lambda})\neq 0$ for all different ${\lambda}\in [\lambda^{(s)}_{i}]$ and all different $j_{\lambda}\in \mathcal{T}$, then select the columns $j_1$, $j_2$, $\ldots$, $j_{\lambda^{(s)}_{i}}$ and all the rows to form a square matrix $\mathbf{M}$ with non-zero diagonal elements, and finally we view the determinant of $\mathbf{M}$ as a non-zero polynomial of $(\mathbf{H}(k_1,j_1), \mathbf{H}(k_2,j_2),\mathbf{H}(k_{\lambda^{(s)}_{i}},j_{\lambda^{(s)}_{i}}))$ \cite{TPCKSC}, so that $\mathbf{M}$ is invertible with high probability by Schwartz-Zippel \cite{DL} in the following.
\begin{lemma}[Schwartz, Zippel lemma \cite{DL}]
\label{le-Sc-Zi}
Let $f\in \mathcal{F}[x_1,x_2,\ldots,x_n]$ be a non-zero polynomial of total degree $d\geq0$ over a field $\mathcal{F}$. Let $\mathcal{S}$ be a finite subset of $\mathcal{F}$ and let $r_1$, $r_2$, $\ldots$, $r_n$ be selected at random independently and uniformly from $\mathcal{S}$. Then, Pr$(f(r_0,r_1,\ldots,r_{n-1})=0)\leq \frac{d}{|\mathcal{S}|}$.
\end{lemma}

From the above introduction,  we now only need to show that we can always find a non-zero path $(\mathbf{H}(k_1,j_1)$, $\mathbf{H}(k_2,j_2)$, $\ldots$, $\mathbf{H}(k_{\lambda^{(s)}_{i}},j_{\lambda^{(s)}_{i}}))$ where $\mathbf{H}(k_{\lambda},j_{\lambda})\neq 0$ for all different $\lambda\in [\lambda^{(s)}_{i}]$ and all different $j_{\lambda}\in \mathcal{T}$. From \eqref{eq-T-cached-packet}, we have that each row of $\mathbf{H}$ has exactly $t$ successive non-zero complex numbers. Recall that $\mathcal{P}^{(s)}_{i}=\{k_1,k_2, \ldots,k_{\lambda^{(s)}_{i}}\}$, $l\in[\lambda^{(s)}_{i}]$. By the connecting assumption between transmitters and users, let
\begin{eqnarray*}
j_1=\min\left(\{k_{1},k_{1}+1, \ldots,k_{1}+L-1\}\bigcap \mathcal{T}\right)\ \text{and}
\end{eqnarray*}  
\begin{eqnarray*}
j_{\lambda}=\!\!\min\!\!\left(\!\left(\!\{k_l,\!k_l+1,\!\ldots,\!k_l\!+L\!-1\!\}\bigcap \mathcal{T}\right)\!\setminus\!\{j_1,\!j_2,\!\ldots,\!j_{\lambda-1}\!\}\!\right)
\end{eqnarray*} for each $\lambda\in [2: \lambda^{(s)}_{i}]$. Clearly, $\mathbf{H}(k_{\lambda},j_{\lambda})\neq 0$ always holds for each $\lambda\in [\lambda^{(s)}_{i}]$ by our placement strategy for the transmitters, i.e.,  \eqref{eq-transmitter-caching-content} and \eqref{eq-T-cached-packet}. Furthermore,  we have that $j_1<j_2<\cdots<j_{\lambda^{(s)}_{i}}$ by our assumption that $k_1<k_2<\cdots<k_{\lambda^{(s)}_{i}}$.

From the above discussion, we have that $\mathbf{H}=\mathbf{H}^{(s)}(\mathcal{P}^{(s)}_{i}, \mathcal{T})$ is full row rank.

\section{Proof of Table \ref{tab-relationship} }
\label{appendix-coro-NDT}
{	
Let us compare the values of $\tau_{\text{XTZ}}$ in Lemma \ref{le-Tao} and $\tau_{\text{new}}$ in Theorem \ref{PC-mapda} according to the values of $\frac{M_\text{U}}{N}$ and $\frac{M_\text{T}}{N}$.
\subsubsection{$\frac{M_\text{U}}{N}+\frac{M_\text{T}}{N}\geq 1$} Clearly we have $\frac{M_\text{U}}{N}+\frac{M_\text{T}}{N} \frac{L}{K}\lceil \frac{K}{L} \rceil\geq 1$ and
\begin{align*}
%\label{eq-case-1}
\tau_{\text{XTZ}} &= \frac{ 1-\frac{M_{\text{U}}}{N}}{ \min\{\frac{M_{\text{T}}}{N}+\frac{M_{\text{U}}}{N}, 1 \} }= \frac{K(1-M_{\text{U}}/N)}{K=\min\{K,\frac{KM_{\text{U}}+M_{\text{T}}L\lceil K/L\rceil}{N}\}}\nonumber\\
& = 1-\frac{M_\text{U}}{N} = \tau_{\text{new}}.
\end{align*}This implies that our scheme also achieves the optimal NDT since the NDT in \cite{XTZ} is optimal.
\subsubsection{ $\frac{M_\text{U}}{N}+\frac{M_\text{T}}{N}< 1$ and  $\frac{M_\text{U}}{N}+\frac{M_\text{T}}{N} \frac{L}{K}\lceil \frac{K}{L} \rceil\geq 1$} According to the value $\frac{M_{\text{T}}L}{N}$ in Lemma \ref{le-Tao}, we have to consider the following two subcases, i.e.,  $\frac{M_\text{T}}{N}=\frac{1}{L}$ and $\frac{M_\text{T}}{N}\in\{\frac{2}{L},\ldots\frac{L-1}{L}, 1\}$. When $\frac{M_\text{T}}{N}=\frac{1}{L}$ we have
\begin{align*}
%\label{eq-case-2}
\tau_{\text{XTZ}} = \left(1-\frac{1}{L}+\frac{1}{ \frac{M_{\text{U}}L}{N} + 1 }\right)\!\!\!\left(1-\frac{M_{\text{U}}}{N}\right)>1-\frac{M_{\text{U}}}{N}=\tau_{\text{new}}.
\end{align*}
When $\frac{M_\text{T}}{N}\in\{\frac{2}{L},\ldots\frac{L-1}{L}, 1\}$, we have
\begin{align*}
%\label{eq-case-3}
\tau_{\text{XTZ}} &=\!\! \frac{ 1-\frac{M_{\text{U}}}{N}}{ \min\{\frac{M_{\text{T}}}{N}+\frac{M_{\text{U}}}{N}, 1 \} }> \frac{K(1-M_{\text{U}}/N)}{K=\min\{K,\frac{KM_{\text{U}}+M_{\text{T}}L\lceil K/L\rceil}{N}\}}\nonumber\\
&=\!\!1-\frac{M_{\text{U}}}{N}=\tau_{\text{new}}.
\end{align*}
\subsubsection{  $\frac{M_\text{U}}{N}+\frac{M_\text{T}}{N} \frac{L}{K}\lceil \frac{K}{L} \rceil<1$}  According to the value $\frac{M_{\text{T}}L}{N}$ in Lemma \ref{le-Tao}, the operation symbol ``$\lceil\rceil$" of  $\tau_{\text{new}}$  in Theorem \ref{PC-mapda}, we have to consider the following subcases.

$\bullet$  When $\frac{M_\text{T}}{N}=\frac{1}{L}$,  we have $\frac{M_\text{T}}{N}=\frac{1}{L}<1 $, then
\begin{align*}
\frac{1}{\frac{M_\text{U}L}{N}+1}-\frac{1}{L}&= \!\! \frac{1}{L}\left(\frac{1}{\frac{M_\text{U}}{N}+\frac{M_\text{T}}{N}}-1\right)< \frac{1}{\frac{M_\text{U}}{N}+\frac{M_\text{T}}{N}}-1
 \\
&\leq \frac{1}{\frac{M_\text{U}}{N}+\frac{M_\text{T}}{N} \frac{L}{K}\lceil \frac{K}{L} \rceil} -1.
	\end{align*}
	Hence, we have $1+ \frac{1}{\frac{M_\text{U}L}{N}+1}-\frac{1}{L} < \frac{1}{\frac{M_\text{U}}{N}+\frac{M_\text{T}}{N}\frac{L}{K}\lceil \frac{K}{L} \rceil }$, and
\begin{align*}
%\label{eq-case-4}
\tau_{\text{new}} &= \frac{ 1-\frac{ M_{\text{U}} }{ N } }{\frac{M_\text{U}}{N}+\frac{M_\text{T}}{N} \frac{L}{K}\lceil \frac{K}{L} \rceil} \\
& > \left(1-\frac{1}{L}+\frac{1}{ \frac{M_{\text{U}}L}{N} + 1 }\right)\left(1-\frac{M_{\text{U}}}{N}\right) = \tau_{\text{XTZ}}.\nonumber
\end{align*} 

$\bullet$  When $\frac{M_\text{T}}{N}\in\{\frac{2}{L},\ldots\frac{L-1}{L}, 1\}$ and $ K\nmid L$ we have
\begin{align*}
%		\label{eq-case-5}
		\tau_{\text{new}} = \frac{ 1-\frac{ M_{\text{U}} }{ N } }{\frac{M_\text{U}}{N}+\frac{M_\text{T}}{N} \frac{L}{K}\lceil \frac{K}{L} \rceil} < \frac{ 1-\frac{ M_{\text{U}} }{ N } }{\frac{M_\text{U}}{N}+\frac{M_\text{T}}{N} } = \tau_{\text{XTZ}}.
\end{align*}

$\bullet$  When $\frac{M_\text{T}}{N}\in\{\frac{2}{L},\ldots\frac{L-1}{L}, 1\}$ and $ K\mid L$ we have
\begin{align*}
%		\label{eq-case-6}
		\tau_{\text{new}} = \frac{ 1-\frac{ M_{\text{U}} }{ N } }{\frac{M_\text{U}}{N}+\frac{M_\text{T}}{N} \frac{L}{K}\lceil \frac{K}{L} \rceil} = \frac{ 1-\frac{ M_{\text{U}} }{ N } }{\frac{M_\text{U}}{N}+\frac{M_\text{T}}{N} } = \tau_{\text{XTZ}}.
	\end{align*} 

}

\end{appendices}

\bibliographystyle{IEEEtran}
\bibliography{reference}

\end{document}